ESTUDIOS / *RESEARCH STUDIES*

# Ventajas comparativas reveladas en disciplinas científicas y tecnológicas en Uruguay

Néstor Gandelman*, Osiris Parcero**, Matilde Pereira***, Flavia Roldan*

*Universidad ORT Uruguay
Correo-e: gandelman@ort.edu.uy ORCID iD: https://orcid.org/0000-0002-4023-8342
Correo-e: roldan@ort.edu.uy ORCID iD: https://orcid.org/0000-0002-6819-2163
** Universitat de les Illes Balears
Correo-e: osirisjorge.parcero@gmail.com ORCID iD: https://orcid.org/0000-0002-6899-7068
***Universidad ORT Uruguay, Universidad de la República y CINVE
Correo-e: matildepereiraelola@gmail.com ORCID iD: https://orcid.org/0000-0001-5096-8691

**Resumen:** En base a información bibliométrica de Scopus para el período 1996-2019, este documento caracteriza la evolución de la producción científica uruguaya y establece las áreas en las cuales el país posee una ventaja comparativa revelada (VCR). Metodológicamente, se propone que se cuenta con una VCR en un área si esta área tiene una participación en la producción científica nacional mayor que la participación del área en la producción científica mundial. La evidencia presentada considera dos mediciones de producción científica (artículos publicados y citas) y tres niveles de agregación en las áreas (una menor con 5 grandes áreas, una más detallada con 27 disciplinas y otra aún más granular con más de 300 desagregaciones). Dentro de Ciencias de la salud se cuenta con VCR en Veterinaria, Enfermería y Medicina. Dentro de Ciencias de la vida se tiene VCR en Ciencias agrícolas y biológicas, Inmunología y microbiología y Bioquímica, genética y bilogía molecular. En Ciencias físicas sólo se tiene VCR en Ciencia medioambiental y en Ciencias Sociales sólo en Economía, econometría y finanzas.

**Palabras clave:** ventajas comparativas reveladas; bibliometría; Scopus; políticas de ciencia y tecnología

**Revealed comparative advantages in scientific and technological disciplines in Uruguay**

**Abstract:** Based on bibliometric information from Scopus for the period 1996-2019, this document characterizes the evolution of Uruguayan scientific production and establishes the areas in which the country has a revealed comparative advantage (RCA). Methodologically, it is proposed that there is a RCA in an area if this area has a greater share in national scientific production than the share of the area in world scientific production. The evidence presented considers two measurements of scientific production (published articles and citations) and three levels of aggregation in the areas (a minor one with 5 large areas, a more detailed one with 27 disciplines and another even more granular with more than 300 disaggregations). Within Health Sciences there is a RCA in Veterinary, Nursing and Medicine. Within Life Sciences there is a RCA in Agricultural and Biological Sciences, Immunology and Microbiology and Biochemistry, Genetics and Molecular Biology. In Physical Sciences there is only a RCA in Environmental Science and in Social Sciences only in Economics, Econometrics and Finance.

## 1. INTRODUCCIÓN

La promoción científica mediante fondos públicos requiere asignar recursos limitados entre una variedad de disciplinas. A modo de ejemplo, el Fondo María Viñas de la Agencia Nacional de Investigación e Innovación (ANII) que financió este proyecto debe evaluar y ordenar los proyectos presentados. Sin embargo, es muy difícil comparar proyectos de distintas áreas. ¿Cómo saber si un proyecto de ciencias sociales es mejor que un proyecto de ciencias de la salud? Muchas veces los integrantes de las comisiones evaluadoras tienen dificultad para entender incluso los títulos de proyectos de áreas que les son lejanas. Entonces, ¿cómo pueden evaluarse los méritos relativos? La realidad es que es una tarea muy compleja. Por ello se establecen mecanismos de preasignación de fondos entre áreas y se confía en los rankings de las comisiones especializadas solo dentro de estas áreas.

Estos mecanismos de preasignación suelen seguir reglas heurísticas muy simples, tales como replicar la proporción de postulaciones por áreas. Las bases del Fondo María Viñas del 2019 establecen que la distribución entre áreas "se realizará en función de la participación de cada uno en la demanda total de pertinentes, medida en cantidad de proyectos." La Comisión Sectorial de Investigación Científica de la Universidad de la República utiliza el mismo criterio. Bianco et al (2014) indican que la asignación de fondos se realiza "de acuerdo al volumen de la demanda por área de conocimientos de manera tal que aquellas áreas que presentan una mayor demanda de propuestas académicamente aprobadas son quienes mayor proporción de recursos reciben en cada convocatoria".

Esto que acabamos de ejemplificar con dos de los programas más importantes del Uruguay, es una norma común de distribución de fondos de investigación en el mundo. En definitiva, es habitual que sea la propia demanda de los investigadores lo que determina las asignaciones por área sin que haya una evaluación efectiva de la conveniencia o efectividad de esta asignación. Naturalmente, esta afirmación no debe tomarse en sentido absoluto. Muchos países han generado planes de desarrollo científico en los que se seleccionan áreas prioritarias o de interés estratégico. Cuando estas políticas están adecuadamente diseñadas, además de definir áreas, designan indicadores de situación, computan una línea de base y establecen objetivos a mediano o largo plazo.

Este documento tiene dos objetivos aplicados a Uruguay pero que pueden reproducirse para otros países. En primer lugar, caracterizar la evolución de la producción científica que ha sido relativamente ignorada por muchos de los vigentes sistemas de evaluación científica en Uruguay y por la literatura nacional. La mayor parte de la evaluación de la actividad científica refiere a los insumos que entran en el proceso. De esta manera, se hace énfasis en los recursos invertidos y su articulación (Bertola y otros, 2005), en la demanda por estos fondos (Robaina y Sutz, 2014) en los recursos humanos en su globalidad, en aspectos etarios y de género de estos recursos humanos (Sclavo y Waiter, 2014 y Goñi, Schenck y Tomassini, 2014) o en la infraestructura disponible (Baptista y otros, 2012). Una notable excepción la constituye el Boletín de Indicadores de Ciencia, Tecnología e Innovación que elabora anualmente ANII donde se distinguen los indicadores de insumos de los indicadores de resultados. A modo ilustrativo del peso relativo de cada uno, en la edición 2017 de este boletín, se destinaron 16 hojas a indicadores de insumos y 2 hojas a indicadores de resultados.

El segundo objetivo de este trabajo es caracterizar los patrones de especialización científica indicando las áreas en las que el país cuenta con ventajas comparativas en su desarrollo y las áreas en que se carece de tales ventajas.

El indicador propuesto es de extensa aplicación en la literatura de comercio internacional. En este estudio extendemos su uso a la evaluación del desempeño de las áreas de producción de conocimiento y proponemos una prueba de significación estadística que no se había usado previamente en las aplicaciones bibliométricas. Metodológicamente, la innovación consiste en el desarrollo de una prueba de significación estadística para la estimación de índices de ventaja comparativa revelada (VCR) para las distintas disciplinas científicas y tecnológicas.

Se tiene una VCR en una disciplina si la participación en la producción nacional de esta disciplina es superior a la participación de la disciplina en el mundo. Esto refleja que, condicional en el estado de desarrollo científico agregado del país, una disciplina con una VCR tiene una proporción mayor en la producción nacional de lo que sería esperable por la participación de la disciplina en la producción mundial. En otras palabras, el país está demostrando mayor capacidad relativa de producción en esa área que el promedio del mundo.

En cuanto a las posibles conclusiones de política científica que se pueda desprender de este estudio, es importante aclarar de modo explícito que la evaluación normativa de los resultados no es única ni determinante. Enfrentados a un área en que Uruguay no tiene una ventaja comparativa revelada las autoridades deberán de decidir el curso de acción. Una opción sería reconocer esta realidad actual y asignarle una participación minoritaria de fondos, o no asignarle directamente ningún fondo. Alternativamente, podría desearse revertir estas desventajas comparativas del presente con una asignación más que proporcional al sector. Naturalmente, esto último sería a expensas de otros sectores en los que sí se tiene ventajas comparativas. De esta manera, los resultados aquí presentados indicarán los costos de la política y los ganadores y perdedores sectoriales podrán visualizarse claramente.

Uruguay posee algunas características diferenciales que lo hacen un caso de interés en cuanto, tiene un relativamente bajo nivel de inversión en investigación y desarrollo que en el 2018 era de 0,45% del PIB, una dependencia absolutamente predominante de fondos públicos para su financiamiento y una historia que muestra escasa diversidad de instituciones dedicadas al desarrollo de ciencia y tecnología en marco de una gobernanza institucional compleja que se señala debe ser reformulada.

En 1849, poco después de la independencia nacional, se funda la Universidad de la República que detentó hasta 1985 el monopolio de la educación universitaria. En este año se autoriza el funcionamiento de la Universidad Católica del Uruguay que se constituye en la primera universidad privada. Sin embargo, no es hasta 1995 donde se aprueba un régimen general (decreto Nº 308/95) que regula la educación terciaria privada. Bajo este régimen se sucede la habilitación de nuevas instituciones, siendo la primera Universidad ORT Uruguay en 1996, seguido de la Universidad de Montevideo en 1997 y la Universidad de la Empresa en 1998. Más recientemente, en 2017, se reconoce a la Universidad CLAEH. En el 2012, se crea por ley una segunda universidad pública con perfil tecnológico bajo el nombre de UTEC. En la actualidad el sistema universitario uruguayo está conformado por dos universidades públicas, cinco universidades privadas y diversos institutos universitarios especializados en una única área y que, por lo tanto, al amparo de la regulación existente, no son denominados universidades. La Universidad de la República continúa siendo la mayor universidad nacional concentrando la mayor parte de los estudiantes, docentes e investigadores. El financiamiento de las universidades privadas proviene de las matrículas estudiantiles y no son beneficiarias de transferencias presupuestales públicas salvo las que correspondan a llamados concursables de corto plazo.

En la segunda mitad del siglo XX las políticas científicas en Uruguay fueron impulsadas desde el Consejo Nacional de Investigaciones Científicas y Técnicas (CONICYT) creado en 1961 y orientadas a la formación de investigadores en el exterior. En 1986, se establece el Programa de Desarrollo de las Ciencias Básicas (PEDECIBA) con foco en biología, química, geociencias, matemática, física e informática. En 1989, la investigación agropecuaria tiene un impulso con la creación del Instituto Nacional de Investigación Agropecuaria (INIA) y en 1991 la Universidad de la República establece la Comisión Sectorial de Investigación Científica (CSIC). (Bortagaray, 2017).

Ya en el siglo XXI se suceden reformas tendientes a jerarquizar la ciencia y tecnología. En el marco de un rediseño institucional se crea en 2006 la Agencia Nacional de Investigación e Innovación (ANII) con el rol de impulsar la innovación, la investigación y la formación estableciendo una variedad de programas concursables abiertos a investigadores de todas las instituciones del medio. Emblema de esto es el desarrollo el Sistema Nacional de Investigadores (SNI) como un sistema de subsidios para investigadores que son evaluados por pares y clasificados en cuatro niveles de acuerdo con su experiencia académica y producción científica. En 2021 el SNI cuenta con 1774 investigadores activos categorizados en seis áreas: Ciencias agrícolas (234), Ciencias médicas y de la salud (222), Ciencias naturales y exactas (611), Ciencias Sociales (371), Humanidades (159) e Ingeniería y Tecnología (177).

En el 2010 se aprueba el Plan Estratégico Nacional de Ciencia y Tecnología en Innovación (PENCTII) que define como áreas tecnológicas a priorizar las TICs, la Biotecnología y "Otros sectores emergentes con potencial e impacto, como la nanotecnología". Se define asimismo priorizar los siguientes seis sectores: 1. Software, Servicios Informáticos Producción Audiovisual, 2. Salud Humana y Animal (incluye Farmacéutica), 3. Producción Agropecuaria y Agroindustrial, 4. Medio Ambiente y Servicios ambientales, 5. Energía, 6. Educación y Desarrollo Social, 7. Logística y Transporte y 8. Turismo. El PENCTII no estableció indicadores medibles ni estableció objetivos concretos. Tampoco generó una línea de base transversal que permitiera medir un punto de partida. A más de diez años de su aprobación no existe ninguna evaluación global de sus efectos. En este marco este documento no debe interpretarse como una evaluación de este.

En la sección 2 se presentan los aspectos metodológicos, seguidos de los resultados en la sección 3 y unas breves conclusiones en la sección 4.

## 2. ASPECTOS METODOLÓGICOS

### 2.1 El índice de VCR

El concepto teórico de ventaja comparativa tiene su origen en la literatura de comercio internacional. Empíricamente existen dificultades en su medición pues requiere establecer los costos de oportunidad del uso de los factores de producción, de manera contrafactual, en ausencia del patrón de comercio existente. Balassa (1965) desarrolla un indicador que, en lugar de medir directamente la ventaja comparativa, señala en qué productos el patrón de comercio "revela" que el país tiene una ventaja comparativa. De esta manera, el índice de ventaja comparativa revelada se usa para determinar de manera indirecta las ventajas comparativas que tiene un país.[1]

Medir ventajas comparativas en la producción de conocimiento científico tiene dificultades similares o mayores a las de medir directamente ventajas comparativas en la producción de bienes y servicios. Esto es, tanto por obtener los costos de oportunidad relevantes como por la medición de los propios factores que entran en la producción científica. Por lo tanto, este documento transita el mismo camino que Balassa en que se le pide a los datos de producción científica que ellos revelen las ventajas comparativas y, así, en forma indirecta obtener los patrones de especialización en ciencia y tecnología. La pregunta por responder es: ¿en qué disciplinas los patrones de producción de conocimiento revelan que un país posee una ventaja comparativa revelada?

La aplicación de este concepto en el análisis bibliométrico admite dos nomenclaturas diferentes. El primer uso se puede encontrar en Frame (1977) quien lo introdujo como Índice de Actividad (Rousseau y Yang, 2012). No obstante, la primera aplicación del índice de ventaja comparativa revelada en cienciometría tal cual fue introducido por Balassa es más reciente y fue presentado con la misma nomenclatura por Lattimore y Revesz (1996).

Definimos el índice de ventaja comparativa revelada (VCR) como el cociente entre la participación de una disciplina en la producción científica de un país y la participación de esta misma disciplina en la producción científica mundial. Un valor superior (inferior) a 1 indica la presencia (ausencia) de una ventaja comparativa revelada en dicha disciplina.

Formalmente, el índice para la disciplina $i$ sería:

$$VCR_i = \frac{x_i/x}{X_i/X}$$

donde $x_i$ es la cantidad de artículos publicados (o citas) en la disciplina $i$ que corresponden a Uruguay, $x$ es la cantidad de artículos publicados (o citas) que corresponden a Uruguay considerando todas las disciplinas del conocimiento, $X_i$ son la cantidad de artículos publicados (o citas) en todo el mundo en la disciplina $i$, y $X$ es la cantidad de artículos (o citas) de todo el mundo en todas las disciplinas.

Reordenando términos, el VCR para la disciplina $i$ puede entenderse como el cociente entre la participación de Uruguay en la investigación mundial de $i$ y la participación de Uruguay en la investigación mundial en general.

$$VCR_i = \frac{x_i/X_i}{x/X}$$

Nótese que, por definición, cada país tendrá en algunas áreas valores de VCR mayores que 1 y en otras áreas menores que 1. No es posible tener ventajas comparativas en todas las áreas simultáneamente y de aquí la diferencia con el concepto de ventajas absolutas.

La especificación de cuál es el indicador adecuado de producción científica es en sí mismo un elemento de debate entre áreas de conocimiento. Hicks (2005 y 2013) argumenta que las Ciencias sociales y las humanidades tienden a tener formas variadas de producción que incorporan los artículos en revistas referidas pero que también incluyen libros y otras formas de producción no indizadas. Ardanche y otros (2014) menciona la existencia de distintas "culturas" de investigación y producción de conocimiento. Esto implica que la comparación de la cantidad absoluta de publicaciones entre Ciencias naturales y Ciencias sociales no es de por sí un buen indicador del desarrollo de cada área.

En cambio, el VCR no está sujeto a esta crítica debido a que compara participaciones relativas. Es posible que las Ciencias sociales en Uruguay tiendan a publicar menos en revistas indizadas que las Ciencias naturales, pero lo mismo sucede con las Ciencias sociales en todo el mundo. De esta manera, la comparación de la participación de las Ciencias sociales en la producción de Uruguay con la participación de las Ciencias sociales en la producción del mundo nos indica sobre el desarrollo relativo de las Ciencias sociales en Uruguay.

En este trabajo la producción científica se aproxima a través de la cantidad de documentos publicados y de citas recibidas por ellos. Cuando el índice se aplica a las citas muestra la ventaja com-parativa en términos del impacto relativo de su producción científica con relación a la producción mundial, mientras que cuando el índice se aplica a las publicaciones muestra la ventaja comparativa de un país específico con relación al mundo, pero en términos de cantidades generadas.

**2.2 Datos**

La fuente de información es SCImago Journal & Country Rank (SJCR). El SJCR es un portal disponible públicamente que incluye las revistas y los indicadores científicos a partir de la información contenida en la base de datos Scopus (Elsevier B.V.) que cubren más de 34.000 títulos de más de 5.000 editores internacionales y 239 países en todo el mundo. La cobertura temporal disponible es 1996-2019. Según Scopus (2020) "*Scopus delivers the most comprehensive overview of the world's research output in the fields of science, technology, medicine, social science, and arts and humanities*".[2]

La producción se clasifica en 5 grupos que llamamos "grandes áreas". Ellas son: Ciencias de la vida, Ciencias de la salud, Ciencias físicas, Ciencias sociales y humanidades, y Multidisciplinarias. Estas se dividen en 27 áreas temáticas principales[3] y más granularmente en 307 disciplinas específicas. Las revistas pueden pertenecer a más de un área temática y en ese caso los artículos allí publicados figurarán en cada área correspondiente.

La base cubre documentos de todas las regiones geográficas siempre que se proporcione un resumen en inglés. Aproximadamente el 22% de los títulos indexados en Scopus se publican en idiomas distintos al inglés (Scopus, 2020).

Consideramos en nuestra medida de producción científica todos los documentos citables en revistas (incluyendo artículos de congresos publicados en revistas). No incluimos series de libros o reseñas de libros, cartas, resúmenes de reuniones de conferencias o fuentes que no sean de series.

En cuanto a impacto de las publicaciones, la base recoge para cada año el número de citas por los documentos publicados en dicho año. Esto significa que un artículo publicado en 2010, en la base de datos utilizada ha acumulado citas para los años 2010 a 2019, y la suma de estas citas se asigna al año 2010.

**2.3. Modelización y construcción de intervalos de confianza**

Al menos para nuestro conocimiento, las aplicaciones al índice VCR suelen basarse en información de corte transversal para una única ventana de tiempo. De esta manera es posible computar el valor puntual del VCR en áreas y países específicos, pero no es posible construir un intervalo de confianza que permita afirmar si el valor encontrado es estadísticamente distinto de 1.

En este documento mostramos cómo es posible realizar esta prueba estadística. Los supuestos básicos son que existe un valor efectivo latente no observado del índice y que las mediciones anuales son realizaciones muestrales del mismo. La serie temporal del VCR se representa mediante la suma de dos componentes

$$VCR_t = VCR_t^* + \varepsilon_t$$

El segundo sumando es un componente aleatorio de media cero, varianza constante y distribución normal. El primer sumando es el nivel del índice que no observamos y suponemos una función del tiempo y de un conjunto de parámetros:

$$VCR_t^* = f(t, \beta)$$

La forma más sencilla de modelar la evolución del nivel de la serie a lo largo del tiempo es suponer que el mismo es constante $VCR_t^* = VCR_{t-1}^* = VCR^*$. De esta manera se puede aplicar una prueba $t$ a la serie $VCR_t$ y testear si en media es estadísticamente distinto de 1.

Un supuesto menos restrictivo, ya que permite modelizar cambios a lo largo del tiempo, es suponer que el VCR sigue una tendencia lineal, $VCR_t^* = \beta_0 + \beta_1 t$. De esta manera la serie temporal responde a:

$$VCR_t = \beta_0 + \beta_1 t + \varepsilon_t$$

y los parámetros pueden ser estimados por mínimos cuadrados ordinarios. Con cada área y país se puede proceder siguiendo estos pasos y así proyectar el valor del VCR para el año deseado T como $VCR_T = \widehat{\beta_0} + \widehat{\beta_1} T$ donde los techos representan a los parámetros estimados. Este valor proyectado es una simple combinación lineal de estimadores con sus errores estándar correspondientes por lo que es posible testear si $VCR_T$ es igual, mayor o menor que cualquier valor de referencia, en nuestro caso 1.

**2.4. Aplicaciones y limitantes del VCR**

Tanto la literatura de comercio internacional como la de cienciometría han discutido ampliamente las ventajas y limitantes del Índice de Ventaja Comparativa Revelada tal como se resume en esta sección.

La facilidad de cálculo del índice VCR, su interpretación sencilla, bajo requerimientos de datos, y su adecuación para la comparación han motivado un amplio uso del índice (Gnidchevenko y Salnikov, 2015; De Benedictis, 2005). El marco original del índice refiere a flujos comerciales (Falkowski, 2017; Grigorovici, 2009; De Benedictis, 2005) pero también se ha llevado a industrias específicas como la forestal (Dieter y Englert, 2007), la fabricación de productos farmacéuticos (Cai, 2018), la agricultura y la alimentación (Jambor y Babu, 2016). Además, el indicador se ha aplicado a otras áreas como el análisis de patentes (Soete y Wyatt, 1983; Zheng y otros, 2011), el comercio electrónico en el sector turístico y el uso de internet (Ruiz Gómez y otros, 2018) y start-ups y capital de riesgo (Guerini y Tenca, 2018).

Tal como se hace en este documento, el índice VCR puede ser calculado para la producción científica utilizando datos bibliométricos para evaluar las ventajas relativas de un país en una disciplina específica. Las medidas bibliométricas, como el índice presentado, contribuyen a investigar y revelar patrones escondidos de tecnología (Daim, 2006; Lee y otros., 2012), a la vez que permiten estudiar la estructura real de los recursos de conocimiento en la nación bajo análisis y en la sociedad como un todo (Radosevic y Yoruk, 2014).

El índice presenta varias ventajas cuando se aplica al análisis de publicaciones. Primero, dado que el conocimiento es un factor crítico para la innovación, el estudio de datos bibliométricos para evaluar el estado de la literatura científica puede contribuir a entender la capacidad científica de una nación (Chuang y otros, 2010; Radosevic y Yoruk, 2014). Aprender acerca de la estructura científica es útil para los hacedores de política ya que informa sobre la distribución sectorial subyacente de recursos asignados a la investigación y es en sí mismo de utilidad en la decisión de nuevas asignaciones (Yang y otros, 2012; Lee y otros, 2012).

Además, dada la fórmula de cálculo, el índice de VCR, tanto aplicado a citas como a documentos, permite una comparación entre naciones que no es dependiente de su tamaño ni de su cantidad absoluta de publicaciones (Mansourzadeh y otros, 2019; Guevara y Mendoza, 2013; Kozlowski y otros, 1999).

Como toda aproximación empírica, el VCR no está exento de inconvenientes. La desventaja más comentada en la literatura refiere a la distribución asimétrica alrededor de su valor neutral (1), con un límite inferior fijo (0) y un límite superior variable entre sectores y a través del tiempo (De Benedictis y Tamberi, 2001 y 2004; Laursen, 2015). El límite superior del VCR depende inversamente de la proporción de las exportaciones del sector en el total de comercio internacional; en nuestro caso, de la proporción de la producción científica mundial que representa cada área. (Schubert y Braun, 1996; Stare y Kejžar, 2014). Algunas transformaciones del VCR, por ejemplo, su transformación logarítmica o la normalización estándar del índice, han sido presentadas en la literatura para reducir el grado de asimetría. (Falkowski, 2017; De Benedictis, 2005; Dalum y otros, 1998). Otra transformación monótona utilizada es la que da lugar al llamado Índice de Especialización Relativa (Rousseau, 2019). Este índice, cuya fórmula es (VCR-1)/(VCR+1), constituye una normalización que preserva el orden estricto del VCR y se utiliza para que los valores del índice permanezcan acotados entre -1 y 1 (Rousseau, 2018; Glänzel, 2000).

Asimismo, un cambio en el índice (decrecimiento o incremento) puede materializarse debido a cambios en la cantidad total de publicaciones mundiales en todas las áreas, que puede no estar relacionado ni con el país ni con el campo de la ciencia bajo estudio (Rousseau, 2018). Esto cobra especial relevancia al momento de analizar políticas públicas específicas en forma aislada, ya que puede ser que estas no sean la causa de un cambio en el índice. Por tanto, Rousseau (2019) tiene una visión negativa del uso del VCR con fines de evaluación de política científicas específicas.

En adición a lo anterior, Mansourzadeh y otros (2019) consideran como una debilidad especial en el uso de datos bibliométricos el potencial problema de asignación de publicaciones. Esto puede deberse tanto a la inclusión de documentos en más de un área temática o por involucrar colaboraciones multinacionales y, por tanto, están incorrectamente asignadas al país correspondiente. Adicionalmente, en el cálculo del VCR se requiere el total mundial de la producción científica, información que sólo se encuentra disponible en algunas bases de datos, como la usada en este estudio.

Entre los trabajos más exhaustivos que utilizan el índice de VCR, el trabajo de Chuang y otros (2010) estima el índice para 26 países que contribuyen hasta el 90% del total de citas y hasta el 86% del total de publicaciones en todo el mundo. El análisis se desarrolla para 24 áreas de la ciencia clasificadas en 6 áreas más abarcativas. Los países son agrupados de acuerdo con el valor del VCR en disciplinas científicas similares con el objetivo de evaluar la capacidad científica de las naciones. Los autores argumentan que los resultados del ejercicio de agrupamiento, que dividen a las naciones en cuatro subgrupos, indican potenciales socios para el desarrollo de colaboraciones en red.

En la misma línea, Radosevic y Yoruk (2014) calculan el índice VCR, tanto para publicaciones como para citas, y desarrollan un análisis de las capacidades científicas de 180 países agrupados en regiones para dos períodos de tiempo (1981-1989 y 2001-2011). Entre los resultados encontrados, se destaca que América del Norte y Europa tienen valores del índice estables que oscilan alrededor del valor uno para todos los campos científicos importantes (con la excepción de ciencias sociales para Europa), mientras que Asia Pacífico se destaca por su mejora en ambos indicadores en las ciencias aplicadas, en contraste con el deterioro de estas medidas para América Latina en el mismo campo. Los resultados además sugieren un importante sesgo hacia las ciencias fundamentales tanto para los países que pertenecían a la ex-URSS como para aquellos de Europa Central y del Este. Por último, los autores concluyen que no ha habido un cambio significativo en la estructura disciplinaria de los sistemas de la ciencia en las regiones del mundo.

## 3. RESULTADOS

### 3.1 Producción científica a lo largo del tiempo en Uruguay

La Figura 1 presenta la evolución de la producción científica anual suavizada con trienios móviles. Dado que contamos con datos desde 1996, el primer trienio móvil disponible es 1996-1998 y este es el origen de las series.

La producción científica mundial ha tenido una notable expansión en las últimas décadas y Uruguay no ha sido la excepción. En 1996, contaba con 272 publicaciones mientras que en 2019 la producción fue de 1.910. La tasa de crecimiento anualizada de Uruguay es de 8,6% mientras que la tasa de crecimiento mundial es de 6,0% y la de América Latina de 9,0% (Figura 1A).

Mayor cantidad de publicaciones debería de redundar en mayor cantidad de citas. Sin embargo, a diferencia de las publicaciones anuales que ex post son invariantes en el tiempo, las citas no lo son. La cantidad de citas de artículos del 2010 será distinta si lo medimos en el 2019 o en el 2020. En el 2019 al observar la cantidad de citas de los artículos publicados en el 2010 estamos observando la cantidad de citas acumuladas en el período 2010-2019. En el 2020 el número de citas de artículos del 2010 aumentará debido a la acumulación de un año adicional. Debido a esto, los artículos publicados más recientemente tenderán a tener menos citas que los artículos publicados previamente, simplemente porque han tenido menos tiempo para acumularlas. Esto se ver reflejado en un patrón temporal de citas con forma de U invertida y por eso para comparar la evolución de citas no es conveniente tomar el último trienio disponible.

El pico de citas anuales para Uruguay se da en el trienio 2010-2012. Los artículos publicados en el trienio 1996-1998 recibieron en promedio 9.037 citas mientras que los publicados en el trienio 2010-2012 recibieron 25.048 citas. Esto implica una tasa de crecimiento anualizada de las citas de 7,9%. En comparación, en el mismo período de tiempo, la tasa de crecimiento anualizada de las citas mundial fue 3,6% y la de América Latina fue de 7,8% (Figura 1B).

En definitiva, considerada en su conjunto la producción uruguaya, medida tanto en cantidad de publicaciones como en citas, ha aumentado a un ritmo superior al del promedio del mundo, pero algo inferior al promedio de América Latina.

La Figura 2 muestra como Uruguay ha tenido una participación creciente en la producción mundial pasando de 0,27 artículos cada 1.000 al comienzo del período a 0,44 artículos cada 1.000 al fin de este.[4] Esto implica que la participación uruguaya en el total de artículos mundiales se multiplicó por 1,6. En relación con América Latina, la producción científica uruguaya se mantuvo relativamente constante en el entorno de 10-11 artículos cada 1.000 publicados. En cuanto a citas, la participación nacional en el contexto mundial y latinoamericano sigue un patrón similar al de los documentos, aunque con un grado mayor de volatilidad.

Si bien existen diferencias entre áreas científicas, la Figura 3 muestra como el patrón de crecimiento en las publicaciones se da para todas las grandes áreas. Tomando las puntas del estudio las publicaciones en Ciencias físicas pasan de 115 a 655, en Ciencias de la salud pasan de 75 a 381, en Ciencias de la vida pasan de 114 a 495 y en Ciencias sociales pasan de 11 a 213. Esto implica tasas de crecimiento promedio anualizadas de 8,7%, 8,1%, 7,3%, 16,1% para Ciencias físicas, Ciencias de la salud, Ciencias de la vida y Ciencias sociales respectivamente (Figura 3.A).

La Figura 3.B muestra que el patrón de citas de los artículos en forma de U invertida es común a todas las áreas, pero existen niveles absolutos muy distintos. Así, las publicaciones en Ciencias de la vida son las que reciben una cantidad mucho mayor de citas, mientras que las publicaciones en Ciencias sociales son las que reciben una cantidad menor.

En las Figura A1-A4 del apéndice online se presentan los resultados de evolución de Uruguay relativa a América Latina y al mundo por grandes áreas tanto en producción absoluta y participación relativa. Se destaca el crecimiento en la participación en documentos mundiales en el período 1996-2019 en las siguientes áreas: Ciencias físicas (de 0,20 a 0,30), Ciencias de la salud (de 0,27 a 0,52), Ciencias de la vida (0.45 a 0.81) y Ciencias sociales (de 0,13 a 0,48), en todos los casos cada 1.000 publicaciones.

Una forma de considerar el impacto promedio de los artículos publicados es considerar la ratio entre citas y cantidad de documentos publicados en un año. Como ya indicamos, los documentos de años más recientes tenderán naturalmente a tener menos citas simplemente porque han tenido menos años de visibilidad. De modo de tener un indicador de la evolución de la calidad de la producción nacional definimos un índice de citación relativa de Uruguay como el cociente entre el patrón de citas anual de Uruguay y el patrón de citas anual del resto del mundo, donde el patrón de citas anual se computa como el cociente entre el total de citas recibidas por el país en un determinado año y la cantidad de documentos publicados. Considerado en su conjunto, existe una tendencia creciente del índice de citación relativo uruguayo en la mayor parte de los años, pero cae al final de la serie (Figura 4). Esta caída en los últimos años puede deberse a la mayor sensibilidad de las estimaciones

**Figura 1.** Producción científica a lo largo del tiempo.

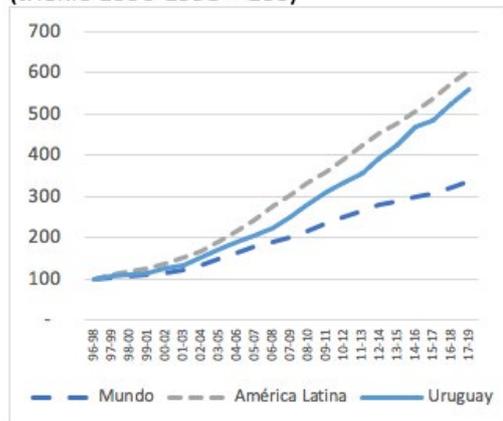
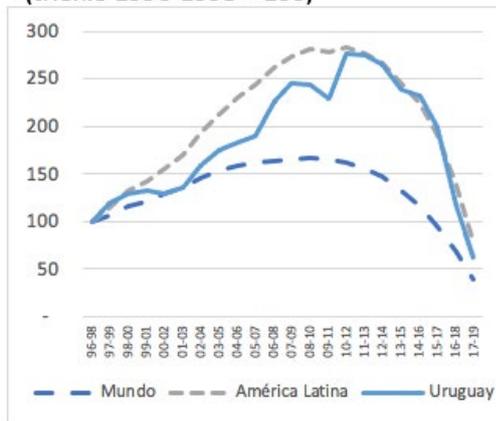

Nota: trienios móviles

**Figura 2.** Participación uruguaya en la producción científica global.

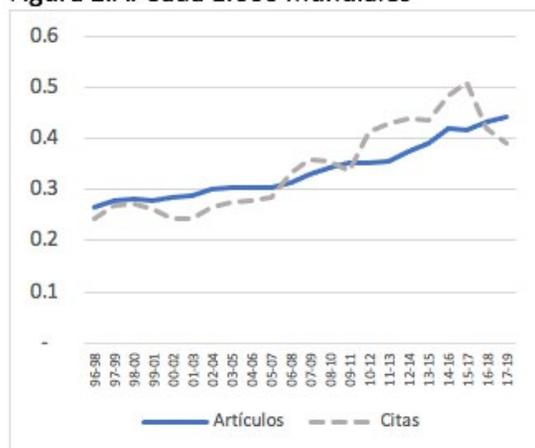
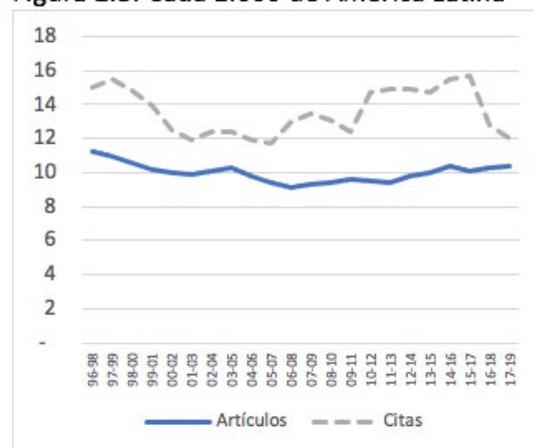

Nota: trienios móviles

**Figura 3.** Producción científica a lo largo del tiempo por grandes áreas.

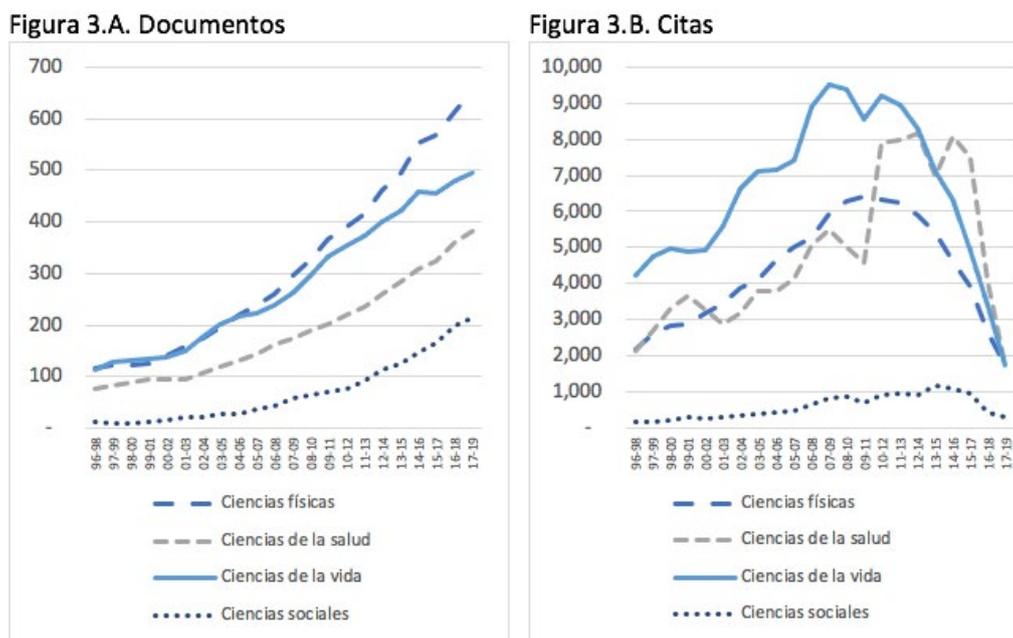

**Figura 4.** Índice de citación relativa de Uruguay respecto al mundo.

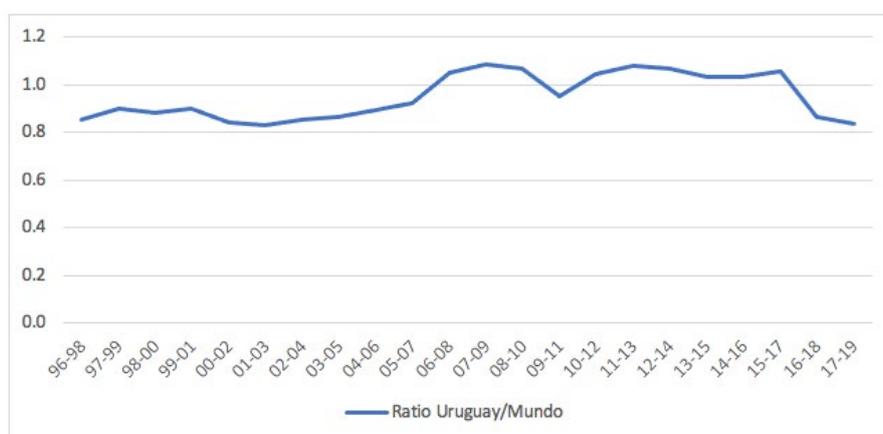

cuando son escasos los años en que se permite acumular citas, especialmente si se considera que los documentos producidos en Uruguay pueden demorar en lograr un grado de exposición que más fácilmente se logra en países de mayor desarrollo científico. El índice de citación relativa agregada es el producto de diferentes áreas que a su vez tienen distintas densidades de citación. En la figura A5 del apéndice se presentan los indicadores de citación relativa por grandes áreas científicas. La fase creciente reportada en el índice general está producida por una mejora de la performance de las ciencias de la salud (que están por arriba del promedio mundial) y de las ciencias sociales (aún por debajo del promedio mundial en la mayor parte del período). Las publicaciones en ciencias físicas muestran una citabilidad por debajo del promedio mundial del área y una tendencia a empeorar su performance según este indicador. Las publicaciones uruguayas de ciencias de la vida tienen una citación promedio inferior a las internacionales, pero una tendencia constante.

### 3.2 VCR en el trienio 2017-2019

La Figura 5 (y Tabla A1 del apéndice online) comienzan la presentación de las ventajas comparativas reveladas, en primera instancia se presentan las estimaciones puntuales del último trienio, sin testear si los resultados son estadísticamente significativos, lo que se hace en la siguiente sección de este trabajo. Las Figuras 5A y 5C muestran como las Ciencias de la vida y las Ciencias de la salud tienen en Uruguay una participación en la producción mayor a la que estas grandes áreas tienen en el mundo. En consecuencia, las figuras 5B y 5D muestra que el VCR para ellas es mayor que 1 indicando que Uruguay cuenta con una VCR en ellas.

En cambio, la producción científica en Ciencias físicas en Uruguay representa una proporción menor que lo que el área representa en el mundo. Esto se traslada en un valor de VCR menor a 1.

En Ciencias sociales se encuentra una situación distinta según se aproxime la producción a través de documentos o citas. Se generan documentos en una proporción muy cerca a la del mundo, pero se recibe una proporción de citas menor a la del mundo. En consonancia, el VCR basado en documentos arroja un valor muy cercano a 1, pero en citas se muestra una desventaja comparativa revelada.

Cada una de estas grandes áreas está compuesta por varias áreas temáticas y ninguna es homogénea a su interior. Dentro de las Ciencias de la salud, es en Veterinaria que se tiene un VCR de más del doble que las demás del área (Figuras 6.A y 7.A). También en Enfermería y Odontología se observa que el índice VCR es mayor a 1. La Figura 7.B muestra que, en las Ciencias de la vida, las de mayor ventaja comparativa revelada son las Ciencias agrícolas y biológicas, seguida de la Inmunología y microbiología y la Bioquímica genética y biología molecular (Figuras 6.B y 7.B). En las Ciencias físicas, la única área temática con un VCR mayor a 1 sería las Ciencias medioambientales (Figuras 6.C y 7.C). En Ciencias Sociales, el mayor valor del índice VCR se encuentra en Economía, econometría y finanzas seguido de las Otras ciencias sociales. Sin embargo, lo anterior se da solo si se aproxima la producción científica por cantidad de documentos (Figura 6.D). Si se aproxima mediante citas, ningún área temática dentro de las Ciencias sociales tiene un valor claramente por arriba de 1 (Figura 7.D).

**Figura 5.** VCR por grandes áreas (trienio 2017-2019).

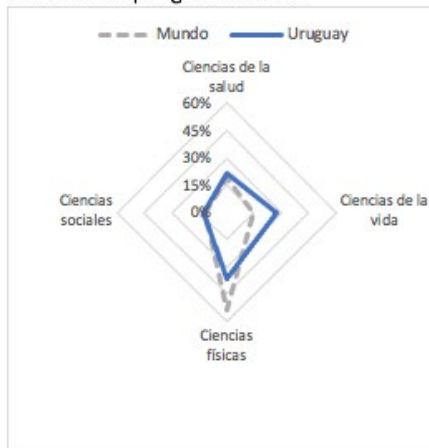
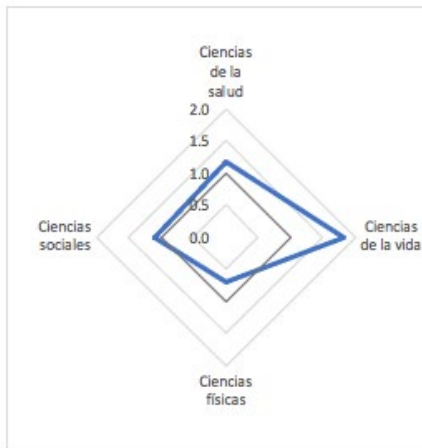
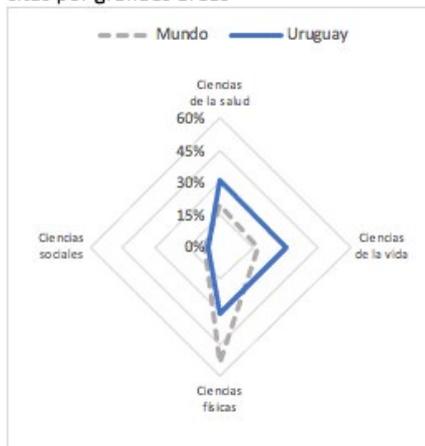
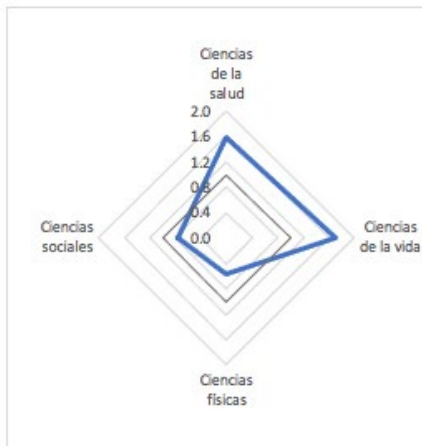

**Figura 6.** VCR por áreas temáticas (basado en documentos, trienio 2017-2019).

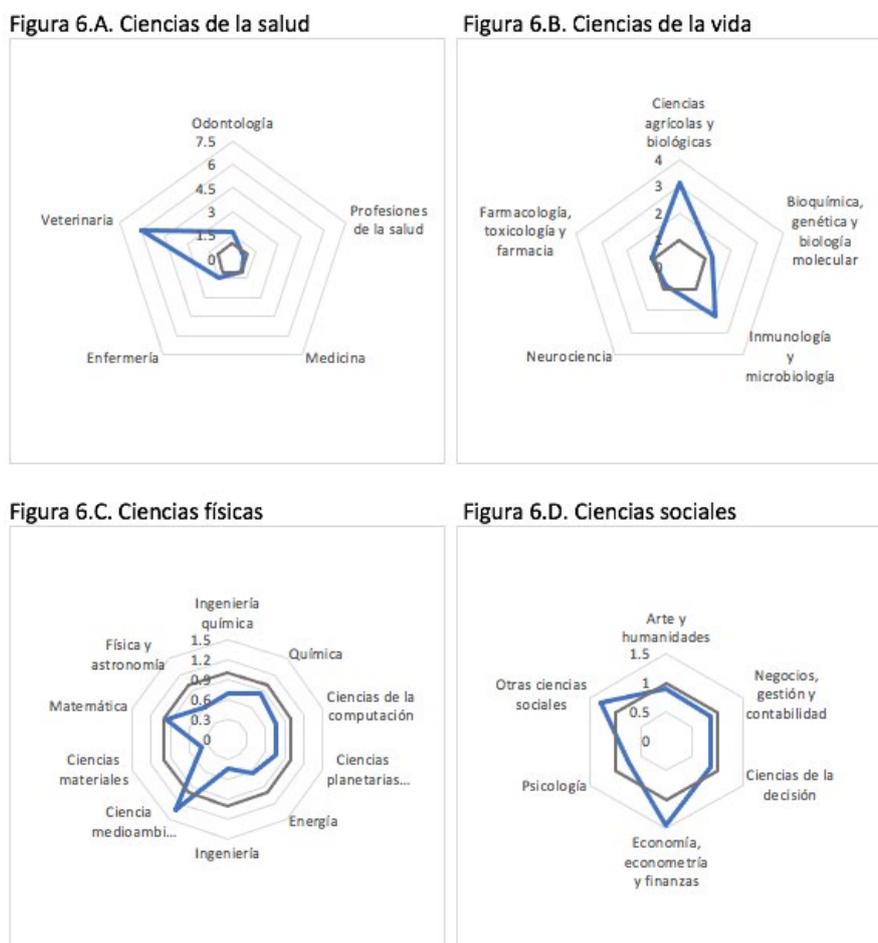

**3.3 Significación estadística del índice de VCR**

La sección anterior presentó los indicadores de ventajas comparativas reveladas basado en la información del último trienio, pero no se testeó si se puede afirmar con certidumbre estadística que los valores sean efectivamente mayores o menores a 1. Para realizar esto es necesario implementar la metodología propuesta en la sección 2.3, que es adicionalmente una prueba de robustez de los resultados.

La Figura 8 y 9 presenta el valor del indicador VCR para cada año junto con la proyección lineal y el intervalo de confianza al 95% de las grandes áreas científicas. De esta manera es fácilmente visualizable para cada punto del tiempo si el VCR es estadísticamente distinto que 1.

Las grandes áreas con una VCR definido y tendencia creciente en la misma son las Ciencias de la salud y las Ciencias de la vida. En Ciencias de la salud la tendencia creciente es especialmente pronunciada cuando la producción científica se aproxima por citas. No sólo se está produciendo una proporción mayor a la esperada en documentos, sino que se está logrando una proporción aún mayor de reconocimiento a este trabajo según citas.

**Figura 7.** VCR por áreas temáticas (basado en citas, trienio 2017-2019).

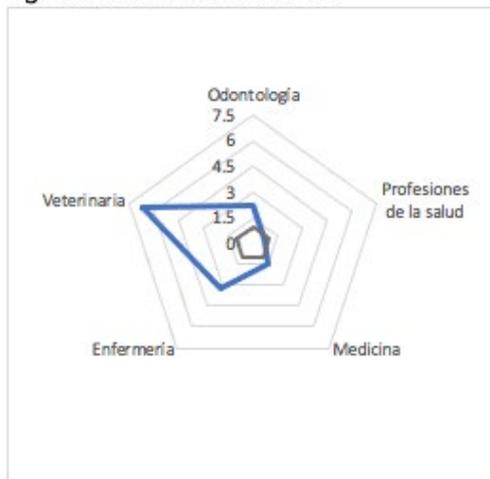
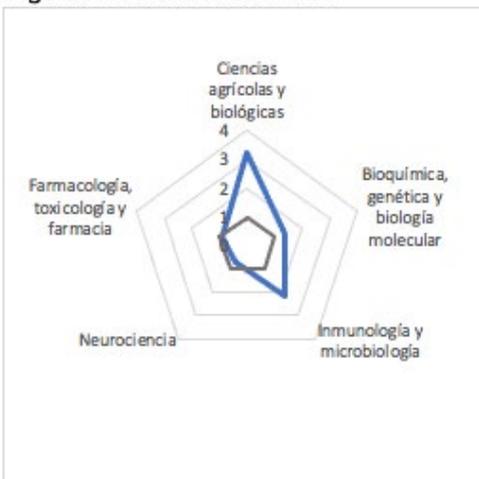
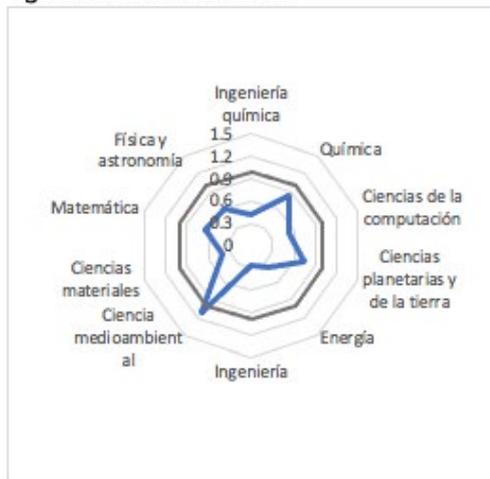
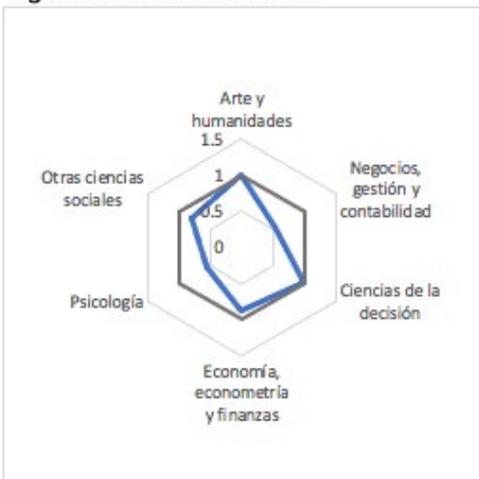

Las Ciencias sociales tienen también una tendencia creciente en su VCR al punto que para el 2019 llegan a obtener un VCR mayor que 1 en documentos. En cuanto a citas, el VCR de Ciencias sociales si bien es también creciente se mantiene indicando una desventaja comparativa relativa.

La situación menos alentadora es en las Ciencias físicas que tienen una desventaja comparativa revelada y una tendencia decreciente en el índice VCR, tanto cuando se mide en cantidad de documentos como cuando se mide en citas.

La gran área definida como Multidisciplinaria (no reportada) muestra mucha mayor volatilidad a lo largo del tiempo, lo que se refleja en un amplio intervalo de confianza y la imposibilidad de establecer estadísticamente si se cuenta con una ventaja o desventaja comparativa revelada.

En resumen, en el extremo positivo, las Ciencias de la vida y Ciencias de la salud tienen una ventaja comparativa revelada estadísticamente significativa (VCR>1) y una tendencia creciente. En un lugar intermedio, las Ciencias sociales tienen una tendencia creciente en su VCR (tanto en documentos como en citas) pero alcanza un valor estadísticamente superior a 1 solo en la medición basada en documentos. En el extremo negativo, las Ciencias físicas tienen una desventaja comparativa revelada (VCR<1) y una tendencia decreciente.

Un análisis con mayor nivel de detalle y desagregación se encuentra en el apéndice online, donde se presentan gráficos similares con la evolución a lo largo del tiempo del VCR para 27 áreas temáticas. En general, el indicador basado en documentos y citas muestra la misma tendencia, pero no en todos los casos.

La Figura 10 presenta la proyección al 2019 del VCR ordenada de mayor a menor junto con sus intervalos de confianza. Se destacan los valores especialmente grandes para Veterinaria y Ciencias agrícolas y biológicas.

La Figura 11 resume las pruebas de VCR. En la parte superior, en verde, se indican las áreas en las que ambos indicadores muestran una ventaja. Estas son: Veterinaria, Ciencias agrícolas y biológicas, Inmunología y microbiología y Enfermería. Seguido, y también pintado de verde, se indican las áreas en las que uno de los indicadores muestra una ventaja comparativa revelada mientras que el otro no rechaza la existencia de dicha ventaja. Estas son: Economía, econometría y finanzas, Bioquímica, genética y biología molecular, Ciencia medioambiental y Medicina.

En la parte inferior, en rojo, figuran las áreas en las que estadísticamente se puede establecer la

**Figura 8.** VCR con intervalo de confianza por grandes áreas (documentos).

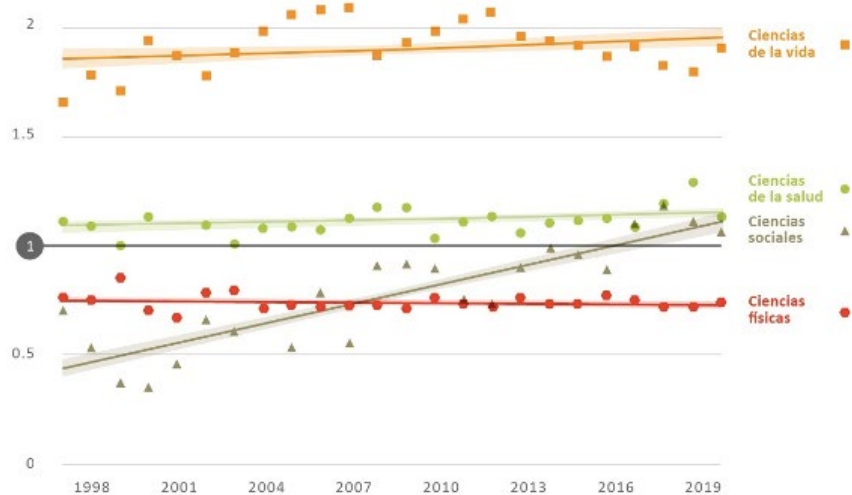

Nota: en sombreado intervalo de confianza al 95%

**Figura 9.** VCR con intervalo de confianza por grandes áreas (citas).

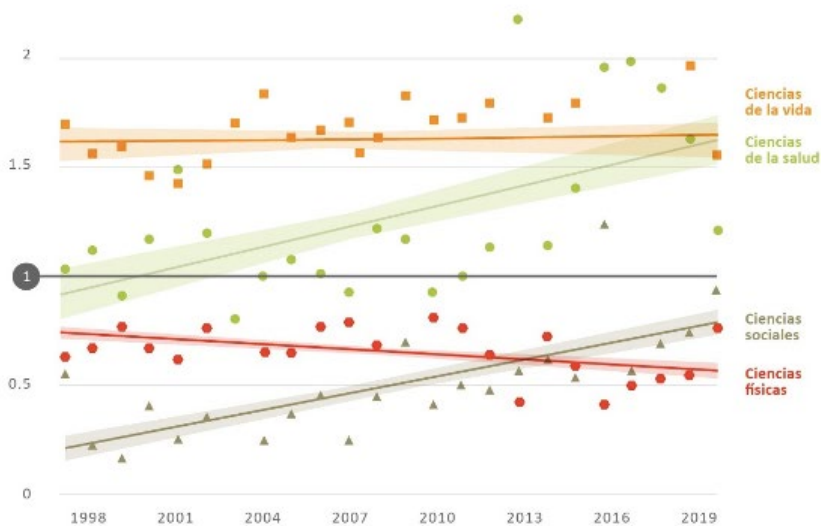

Nota: en sombreado intervalo de confianza al 95%

existencia de una desventaja relativa tanto según documentos publicados como citas recibidas. Estas son: Ciencias de la computación, Ciencias planetarias y de la tierra, Negocios, gestión y contabilidad, Psicología, Ingeniería, Energía, Profesiones de la salud, Física y astronomía, Ingeniería química y Ciencias materiales. Por encima de ellas, pero también en rojo, figuran las áreas en las que un indicador señala estadísticamente la existencia de una desventaja relativa y el otro no rechaza esta desventaja. Estas son las áreas de Matemática y Química.

Finalmente, existe un conjunto de áreas en que tanto el índice VCR basado en documentos como el basado en citas es inconcluso sobre la existencia de ventajas o desventajas reveladas. Estas son: Farmacología, toxicología y farmacia, Odontología, Neurociencia, Multidisciplinarias, Ciencias de la decisión y Arte y humanidades. Las Otras ciencias sociales arrojan resultados contradictorios en cuanto a su VCR según se refiera al indicador de documentos o de citas.

El análisis desarrollado en este documento puede llevarse a niveles aún más desagregados. Para el lector interesado se deja en la Tabla A2 del apéndice

**Figura 10.** Ventajas comparativas reveladas por áreas temáticas (2019).

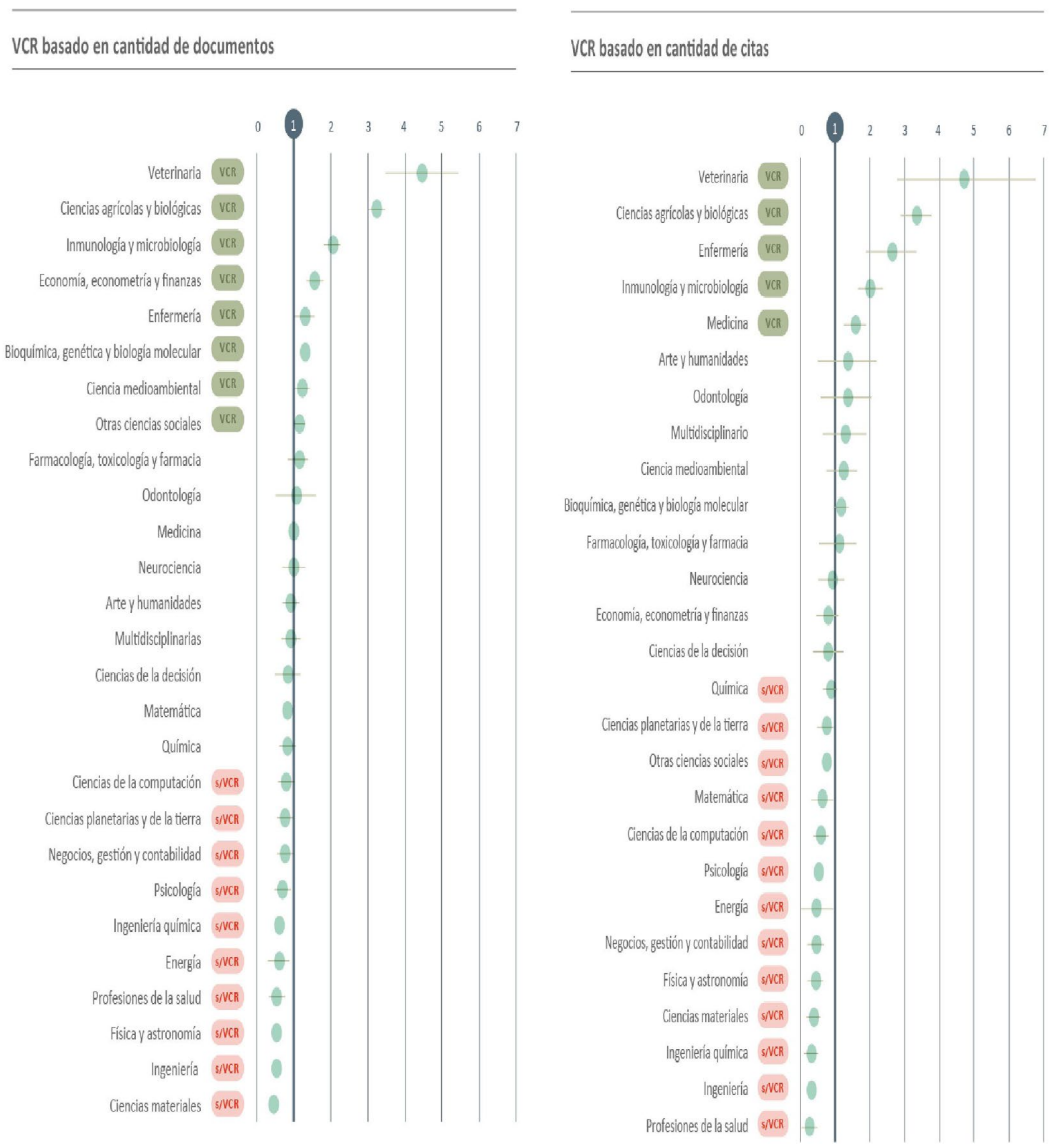

Nota: las líneas horizontales indican el intervalo de confianza al 95%, cuando esta línea no corta el 1 significa que el valor puntual estimado (círculo) es estadísticamente distinto de 1 y se tiene una VCR (si es mayor) o se carece de una VCR (si es menor).

**Figura 11.** Resumen VCR en documentos y citas (2019).

| Áreas | Ventaja comparativa revelada | VCR inconcluso | Desventaja comparativa revelada |
|---|:---:|:---:|:---:|
| Veterinaria | ●○ | | |
| Ciencias agrícolas y biológicas | ●○ | | |
| Inmunología y microbiología | ●○ | | |
| Enfermería | ●○ | | |
| Economía, econometría y finanzas | ● | ○ | |
| Bioquímica, genética y biología molecular | ● | ○ | |
| Ciencia medioambiental | ● | ○ | |
| Medicina | ○ | ● | |
| Farmacología, toxicología y farmacia | | ●○ | |
| Odontología | | ●○ | |
| Neurociencia | | ●○ | |
| Multidisciplinarias | | ●○ | |
| Ciencias de la decisión | | ●○ | |
| Arte y humanidades | | ●○ | |
| Otras ciencias sociales | ● | | ○ |
| Matemática | | ● | ○ |
| Química | | ● | ○ |
| Ciencias de la computación | | | ●○ |
| Ciencias planetarias y de la tierra | | | ●○ |
| Negocios, gestión y contabilidad | | | ●○ |
| Psicología | | | ●○ |
| Ingeniería | | | ●○ |
| Energía | | | ●○ |
| Profesiones de la salud | | | ●○ |
| Física y astronomía | | | ●○ |
| Ingeniería química | | | ●○ |
| Ciencias materiales | | | ●○ |

● Documentos  ○ Citas

online la reproducción de las estimaciones presentadas en las secciones 3c (VCR en el último trienio) y 3d (VCR proyectado 2019 con nivel de significación estadísticas) para más de 300 disciplinas.

## 4. CONCLUSIONES

La asignación de recursos entre áreas disciplinares es una tarea compleja para la cual suele acudirse a reglas de preasignación muchas veces determinadas por la demanda de fondos de parte de los investigadores o de las instituciones. Sin embargo, en la consideración estratégica de la asignación de recursos es necesario también dar relevancia a las dimensiones que aporta este trabajo o similares.

Este trabajo aporta evidencia para constituirse en un insumo de relevancia al momento de evaluar el desempeño sectorial científico, y al momento de diseñar y reconsiderar estrategias de apoyo a la producción científica. El foco en los resultados empíricos está en Uruguay, pero las líneas de análisis son implementables por investigadores con interés en otros países y que pueden encontrar en este documento una guía factible de trabajo.

Metodológicamente, este trabajo realiza una contribución a la literatura mostrando una forma sencilla de computar intervalos de confianza y tes-tear de esta manera la significación estadística de los patrones de especialización científica.

En concreto, en este trabajo hemos i) caracterizado la evolución de la producción científica del Uruguay, y (ii) caracterizado sus patrones de especialización científica, indicando las áreas en las que el país revela tener ventajas comparativas. Esto se realizó en base a información bibliométrica de acceso público para el período 1996-2019.

Antes de repasar los resultados concretos de este estudio debemos remarcar las recomendaciones que se desprenden.

Al igual que en buena parte del mundo, Uruguay tiene una cultura establecida de evaluación de investigadores en función de su producción científica. Sin embargo, esto no se hace a nivel agregado de áreas científicas o de la totalidad de la comunidad científica. El primer mensaje de nuestro documento es que, si bien debemos continuar monitoreando los insumos que entran en el proceso de producción científica, debemos incorporar la medición de los resultados con alguna aproximación a la producción.

En segundo lugar, reconocemos que la medición de la producción científica es en sí mismo un tema de debate, especialmente entre áreas de conocimiento. Nuestra aproximación es bibliométrica. No es la única posible y tampoco pretende ser una visión reduccionista. En cambio, tiene la ventaja de permitir la determinación de parámetros objetivos y que pueden seguirse consistentemente en el tiempo y a través de disciplinas. Todos los indicadores descriptivos de una sociedad tienen limitaciones. La tasa de desempleo es la ratio entre los desempleados y la población económicamente activa. Según las recomendaciones de las conferencias internacionales de estadígrafos del trabajo toda persona que haya trabajado al menos una hora en la semana anterior se considera empleada. Es posible que muchos encuentren está definición como restrictiva, sin embargo, los institutos de estadística siguen esta metodología estándar y la repiten consistentemente mes a mes. Los estadísticos deben ser mirados en función de lo que dicen y comprendiendo sus limitaciones. Los presentados en este trabajo también. Los problemas reseñados y el debate sobre qué es producción no puede ser una invitación a no medir. Por el contrario, refuerza la necesidad de profundizar este debate y la política pública debería lograr, a nivel nacional, definir indicadores de consenso.

En tercer lugar, con indicadores de producción y de insumos estandarizados podremos avanzar en los tan relevantes temas de productividad y eficiencia de los cuales hoy tenemos solo evidencia precaria.

En cuarto lugar, el monitoreo de la marcha de una disciplina no puede hacerse en términos aislados de la marcha de la disciplina en el mundo. El conocimiento científico es esencialmente internacional y, por lo tanto, todos los indicadores de resultados deberían considerarse en función de un marco de referencia acorde.

Finalmente, en cuanto a las posibles conclusiones de política científica sectorial debemos reforzar lo indicado en la introducción: la evaluación normativa de nuestros resultados no es única ni determinante. Hemos presentado áreas en las que Uruguay posee y carece de ventajas comparativas reveladas, pero son las autoridades políticas las que deben decidir los planes de acción. Una alternativa sería focalizar la asignación de fondos en áreas en donde se tienen ventajas comparativas y relegar a aquellas en que no. En oposición, las autoridades pueden querer priorizar presupuestalmente las áreas consideradas estratégicas donde se está hoy en una situación desventajosa. Este tipo de decisiones no se pueden desprender de trabajos como el nuestro, lo máximo a lo que podemos aspirar es a ofrecer insumos. La decisión de política científica recae indefectiblemente en los hacedores de política en su rol de autoridades y no de científicos.

Desde el punto legal-regulatorio, en Uruguay sigue vigente el Plan Estratégico Nacional de Ciencia y Tecnología en Innovación (PENCTII) aprobado en el 2010. En este plan se definieron áreas tecnológicas y sectores a priorizar. Sin embargo, no se estableció indicadores medibles, ni la línea de base ni los objetivos a alcanzar. Esto hace muy difícil concebir una evaluación global de sus efectos. Nuestro documento no lo es. Aspiramos a ser un insumo de utilidad para un próximo plan que esperamos recoja la necesidad de definir indicadores de resultados objetivos, observables por la comunidad y comparables internacionalmente. Este plan deberá establecer la situación de partida y el destino al que se pretende llegar en estos indicadores.

A modo de resumen, los principales resultados encontrados son:

- La producción científica uruguaya, la producción de América Latina y la producción mundial aumentaron considerablemente en el período con tasas promedio anualizadas de 8,6%, 9,0% y 6,0% respetivamente.

- En función de esto, aumentó la participación uruguaya en la producción científica mundial y se mantuvo constante en relación con América Latina.

- En cuanto a grandes áreas, tanto si se calcula el VCR en función de la cantidad de artículos publicados como en cantidad de las citas recibidas, el país tiene VCR en Ciencias de la vida y Ciencias de la salud. En cambio, carece de VCR en Ciencias físicas. La estimación para Ciencias sociales arroja valores contradictorios según se aproxime la producción por cantidad de artículos o citas.

- Existe heterogeneidad al interior de estas grandes áreas.

- Tanto en la medición basada en artículos como en citas Uruguay cuenta con una VCR en Veterinaria, Ciencias agrícolas y biológicas, Inmunología y microbiología y Enfermería.

- Cuenta con VCR en al menos una de las dos mediciones (y la otra es estadísticamente no significativa) en: Economía, econometría y finanzas, Bioquímica, genética y bilogía molecular y en Ciencia medioambiental y Medicina.

- Carece de VCR en: Ciencias de la computación, Ciencias planetarias y de la tierra, Negocios, gestión y contabilidad, Psicología, Ingeniería, Energía, Profesiones de la salud, Física y astronomía, Ingeniería química, Ciencias materiales.

- Los resultados son menos claros en Otras ciencias sociales (resultados estadísticamente significativos pero contradictorios del VCR por documentos y el VCR por citas), Matemática y Química (VCR estadísticamente menor a 1 en una mediciones y no estadísticamente distinto de 1 en la otra).

- En el resto de las disciplinas no se puede establecer si los resultados son estadísticamente mayores o menores a 1. Esto incluye: Farmacología, toxicología y farmacia, Odontología, Neurociencia, Multidisciplinarias, Ciencias de la decisión y Arte y humanidades

En la Tabla A2 del apéndice online se deja a disposición de los lectores los resultados de este análisis a nivel mucho más fino según una desagregación en más de 300 disciplinas. Para hacedores de política especializados, como ser autoridades universitarias, y que quieren considerar grupos de trabajo más precisos, los datos de este apéndice online deberían de ser de utilidad.

## 5. NOTAS

1. Ver por ejemplo, Fertö y Hubbard (2003) para Hungría, Amoroso et al (2011) para México y Depetris y otros (2009) para Argentina y Uruguay.
2. Existen bases alternativas de producción científica, pero muchas de ellas con foco en áreas del conocimiento (por ejemplo, Pubmed en la literatura biomédica) o zonas geográficas (por ejemplo, Scielo con presencia de países de América Latina y Sudáfrica). Web of Science es una alternativa a Scimago con similar amplitud de áreas y zonas consideradas y, por lo tanto, puede ser considerado para un ejercicio como el de este documento. Desde nuestro punto de vista, Scimago presenta dos ventajas respecto a Web of Science. Primero, tiene una cobertura de revistas mayor lo que para un estudio en un país como Uruguay que no se encuentra en la elite científica es importante para poder capturar de mejor manera sus publicaciones. Segundo, se puede acceder a los datos en forma gratuita y libre lo que facilita la reproductibilidad, replicación y seguimiento de los resultados.
3. Arte y humanidades, Bioquímica, genética y biología molecular, Ciencias medioambientales, Ciencias agrícolas y biológicas, Ciencias de la computación, Ciencias de la decisión, Ciencias materiales, Ciencias planetarias y de la tierra, Economía, econometría y finanzas, Energía, Enfermería, Farmacología, toxicología y farmacia, Física y astronomía, Ingeniería, Ingeniería química, Inmunología y microbiología, Matemática, Medicina, Multidisciplinario, Negocios, gestión y contabilidad, Neurociencia, Odontología, Otras ciencias sociales, Profesiones de la salud, Psicología, Química y Veterinaria.
4. Como marco de comparación, vale la pena notar que la población de Uruguay representa 0.45 de cada 1.000 habitantes del planeta.

## 6. REFERENCIAS

**APÉNDICE ONLINE**

**Figura A1.** Evolución de la producción científica en Ciencias físicas

Panel A. Evolución de las publicaciones

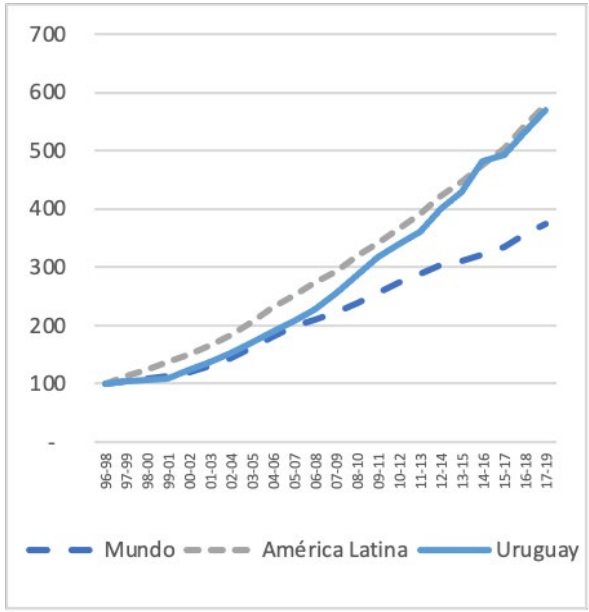

Panel B. Evolución de las citas

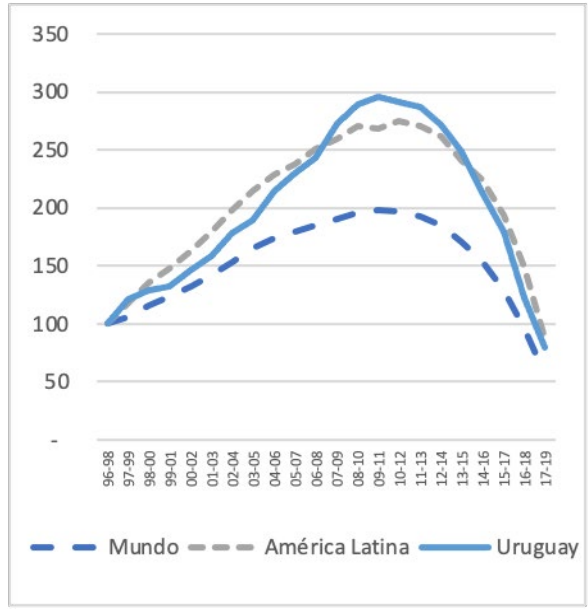

Panel C. Participación de los documentos y citas uruguayas cada 1.000 mundiales

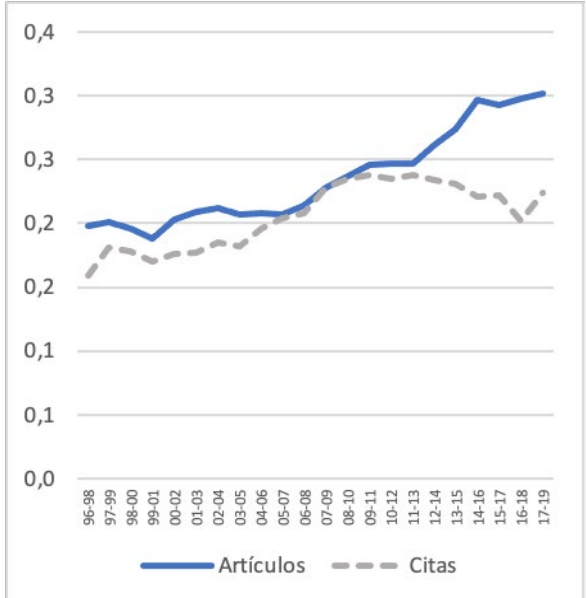

Panel D. Participación de los documentos y citas uruguayas cada 1.000 latinoamericanas

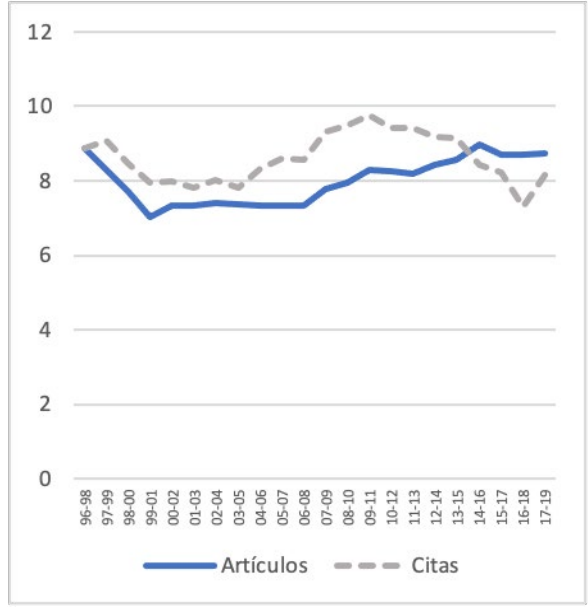

Nota: trienios móviles

**Figura A2.** Evolución de la producción científica en Ciencias de la salud

Panel A. Evolución de las publicaciones

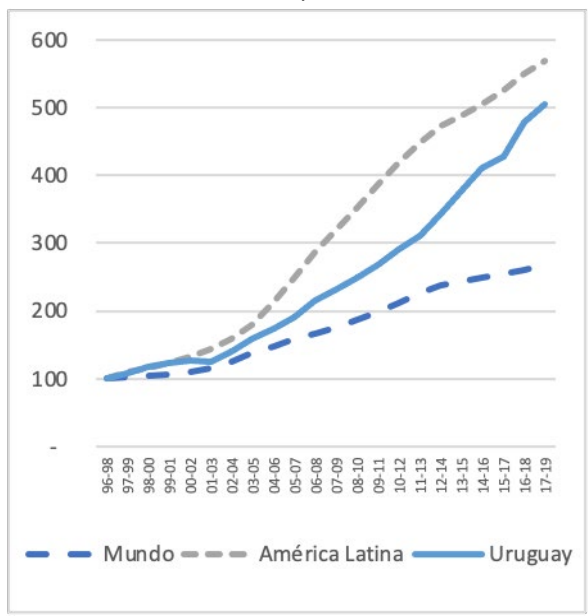

Panel B. Evolución de las citas

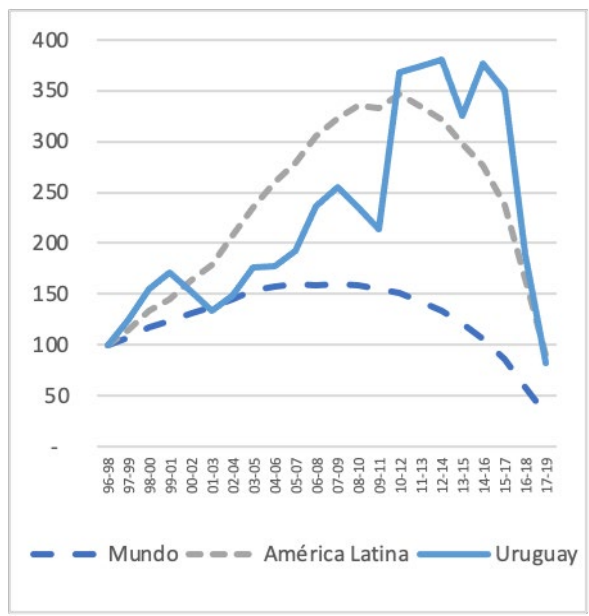

Panel C. Participación de los documentos y citas uruguayas cada 1.000 mundiales

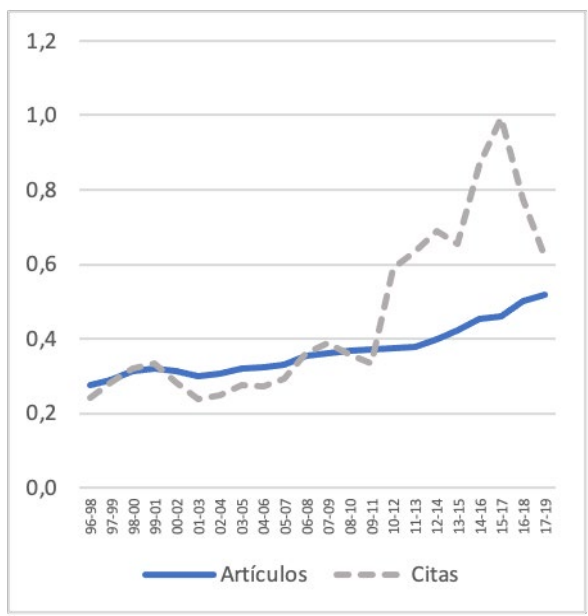

Panel D. Participación de los documentos y citas uruguayas cada 1.000 latinoamericanas

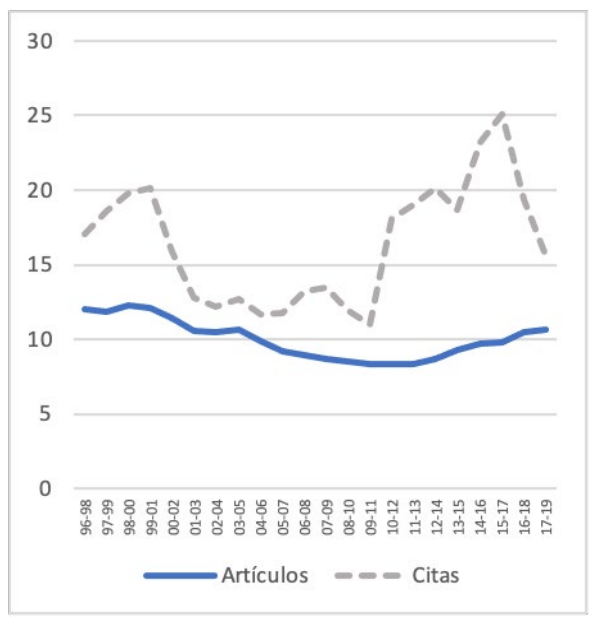

Nota: trienios móviles

**Figura A3.** Evolución de la producción científica en Ciencias de la vida

Panel A. Evolución de las publicaciones

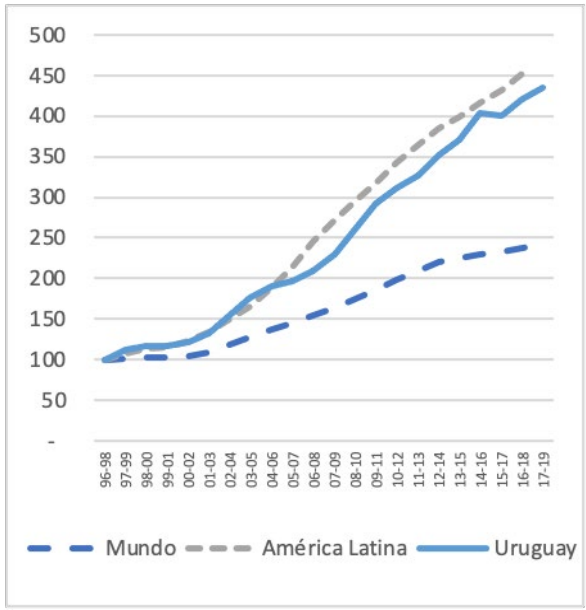

Panel B. Evolución de las citas

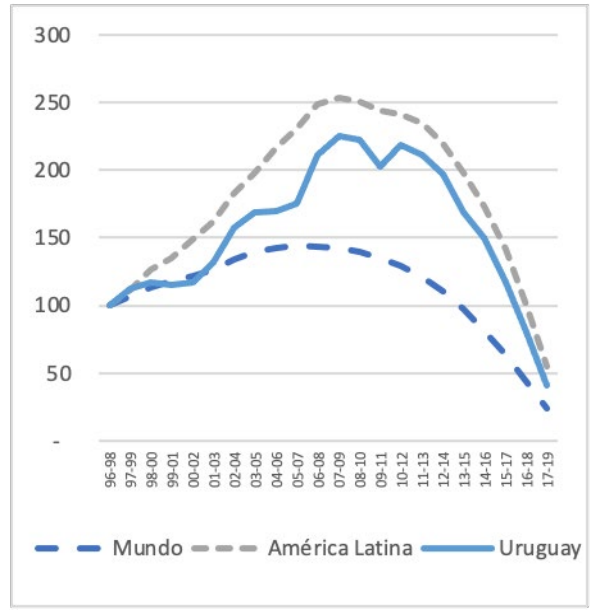

Panel C. Participación de los documentos y citas uruguayas cada 1.000 mundiales

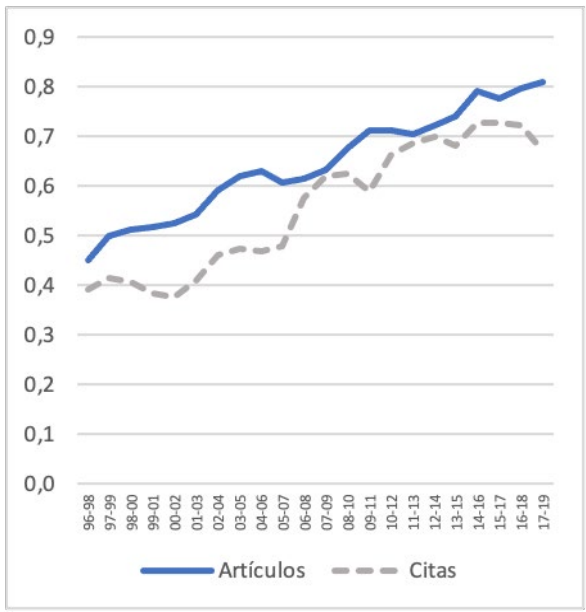

Panel D. Participación de los documentos y citas uruguayas cada 1.000 latinoamericanas

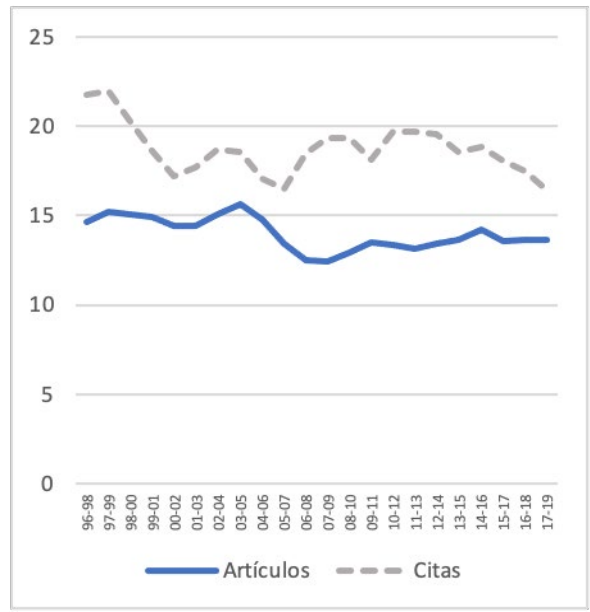

Nota: trienios móviles

**Figura A4.** Evolución de la producción científica en Ciencias sociales

Panel A. Evolución de las publicaciones

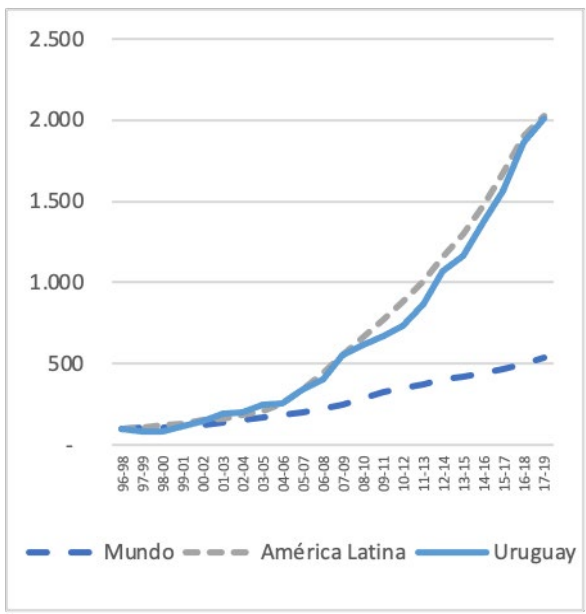

Panel B. Evolución de las citas

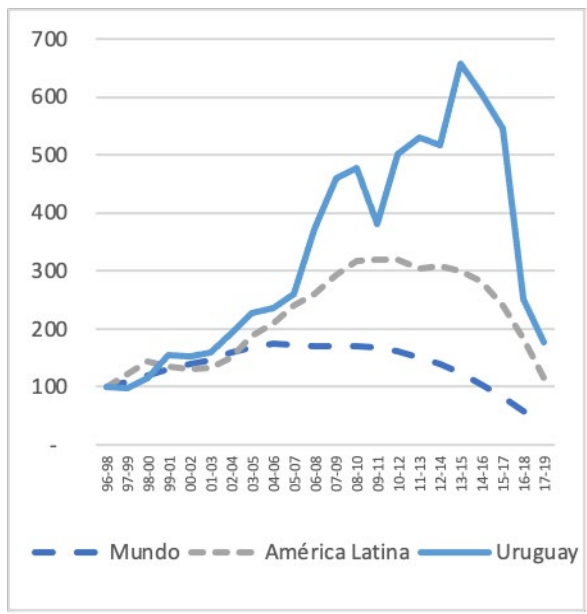

Panel C. Participación de los documentos y citas uruguayas cada 1.000 mundiales

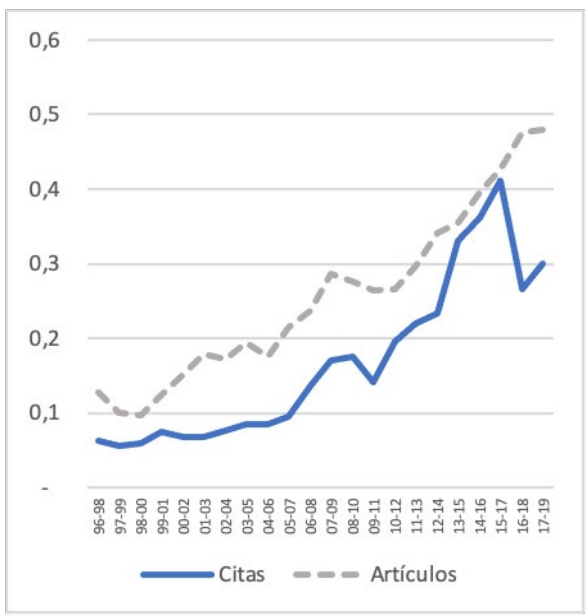

Panel D. Participación de los documentos y citas uruguayas cada 1.000 latinoamericanas

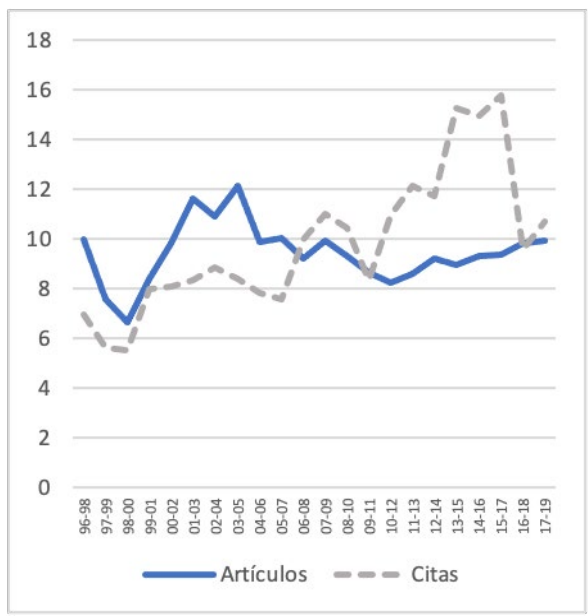

Nota: trienios móviles

**Figura A5.** Índice de citación relativa de Uruguay respecto al mundo por grandes áreas

Panel A. Ciencias físicas

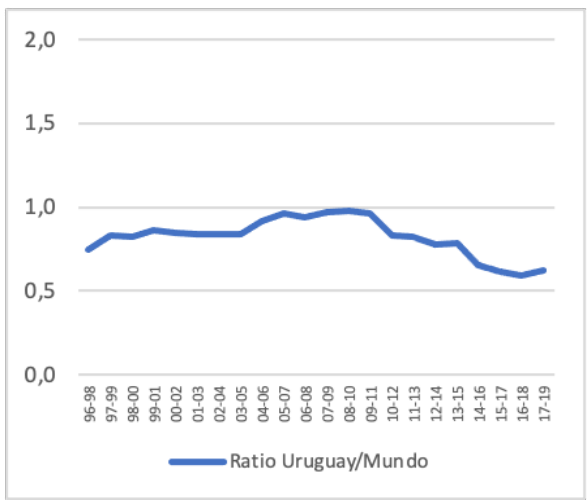

Panel B. Ciencias de la salud

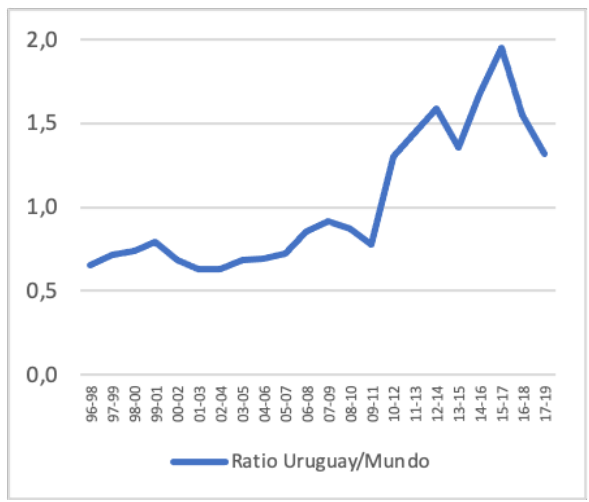

Panel C. Ciencias de la vida

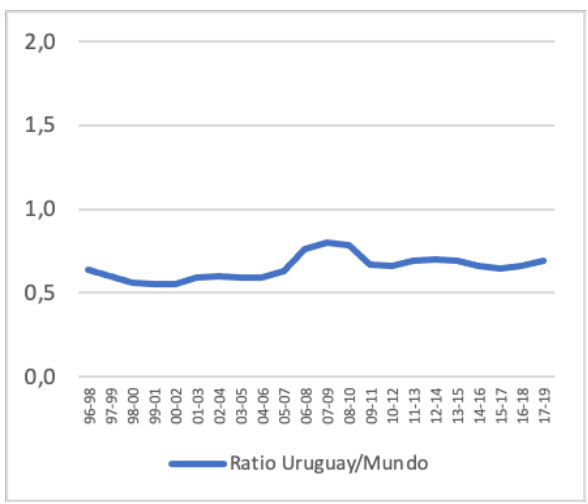

Panel D. Ciencias sociales

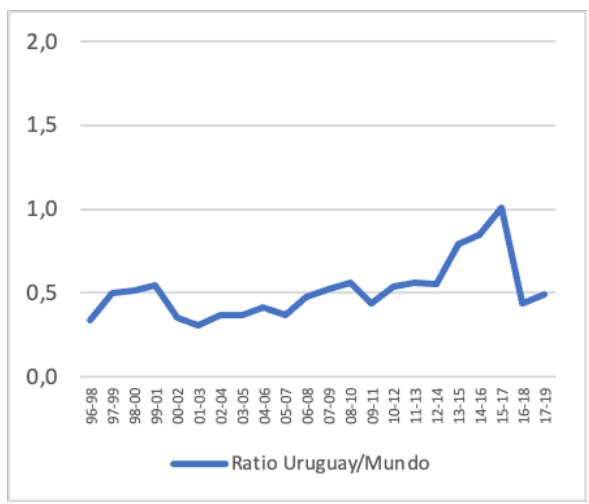

Nota: trienios móviles

**Figura A6:** VCR por áreas temáticas con intervalo de confianza en Ciencias físicas

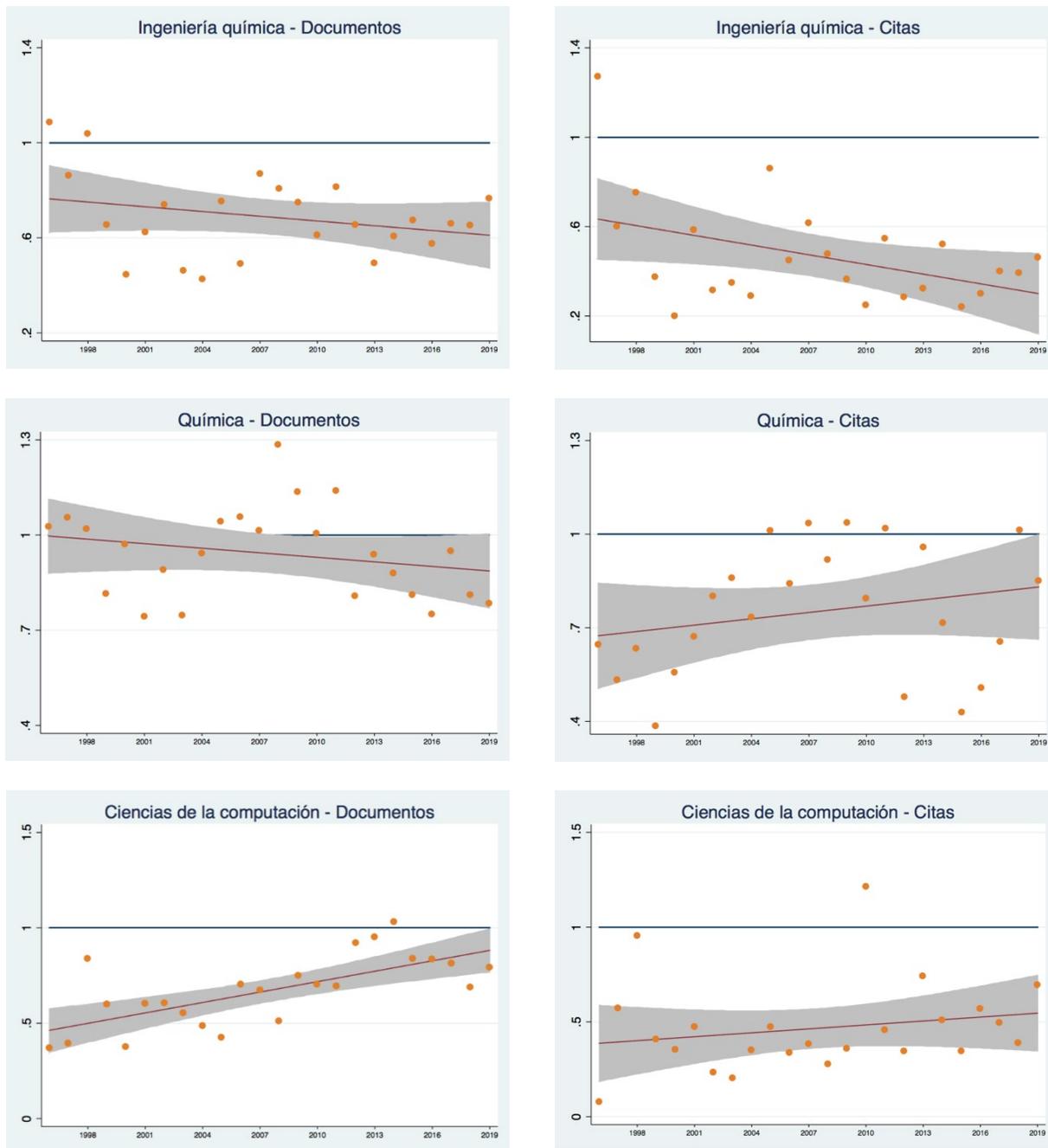

Nota: en gris intervalo de confianza al 95%

**Figura A6 (cont):** VCR por áreas temáticas con intervalo de confianza en Ciencias físicas

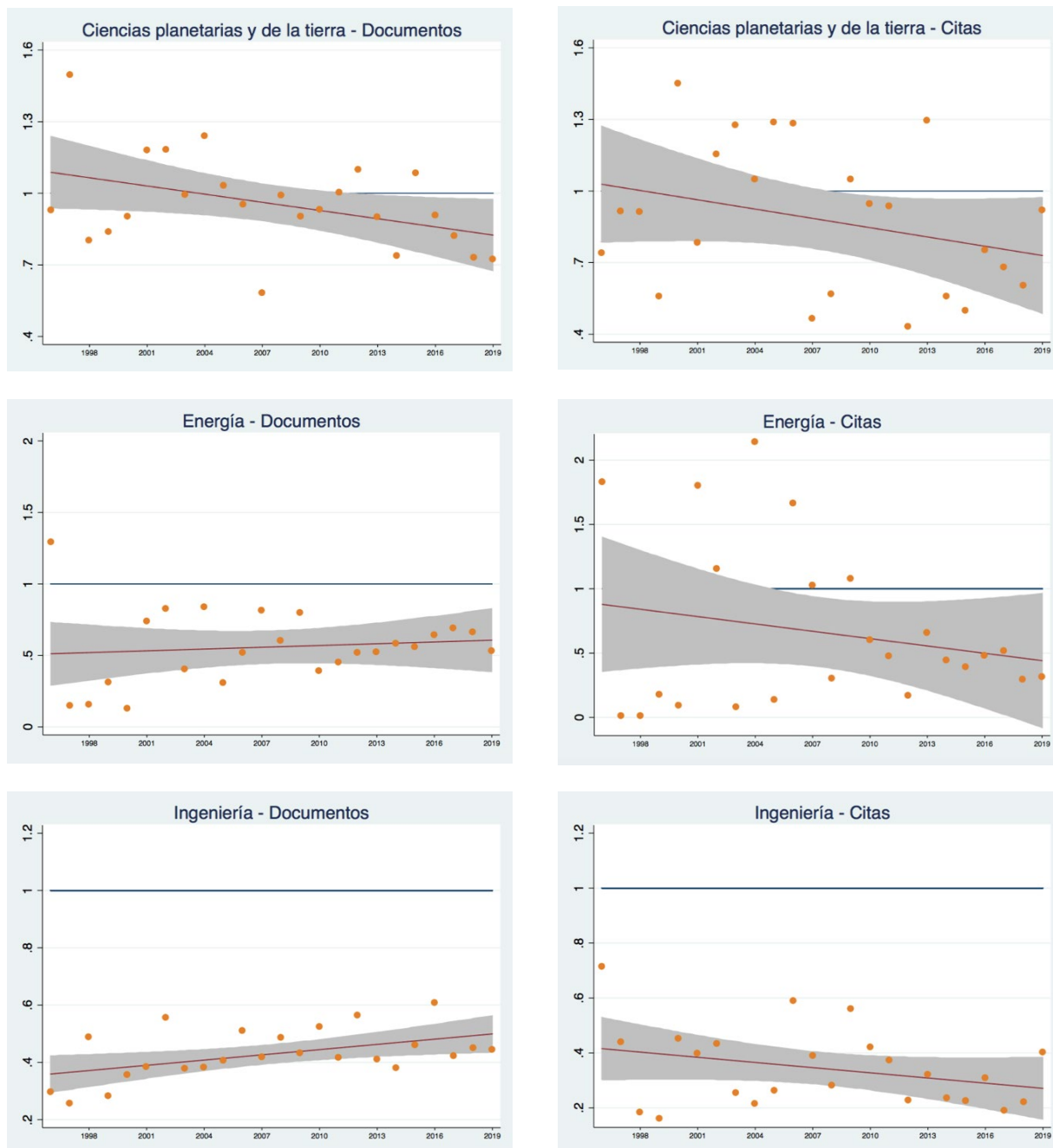

Nota: en gris intervalo de confianza al 95%

**Figura A6 (cont):** VCR por áreas temáticas con intervalo de confianza en Ciencias físicas

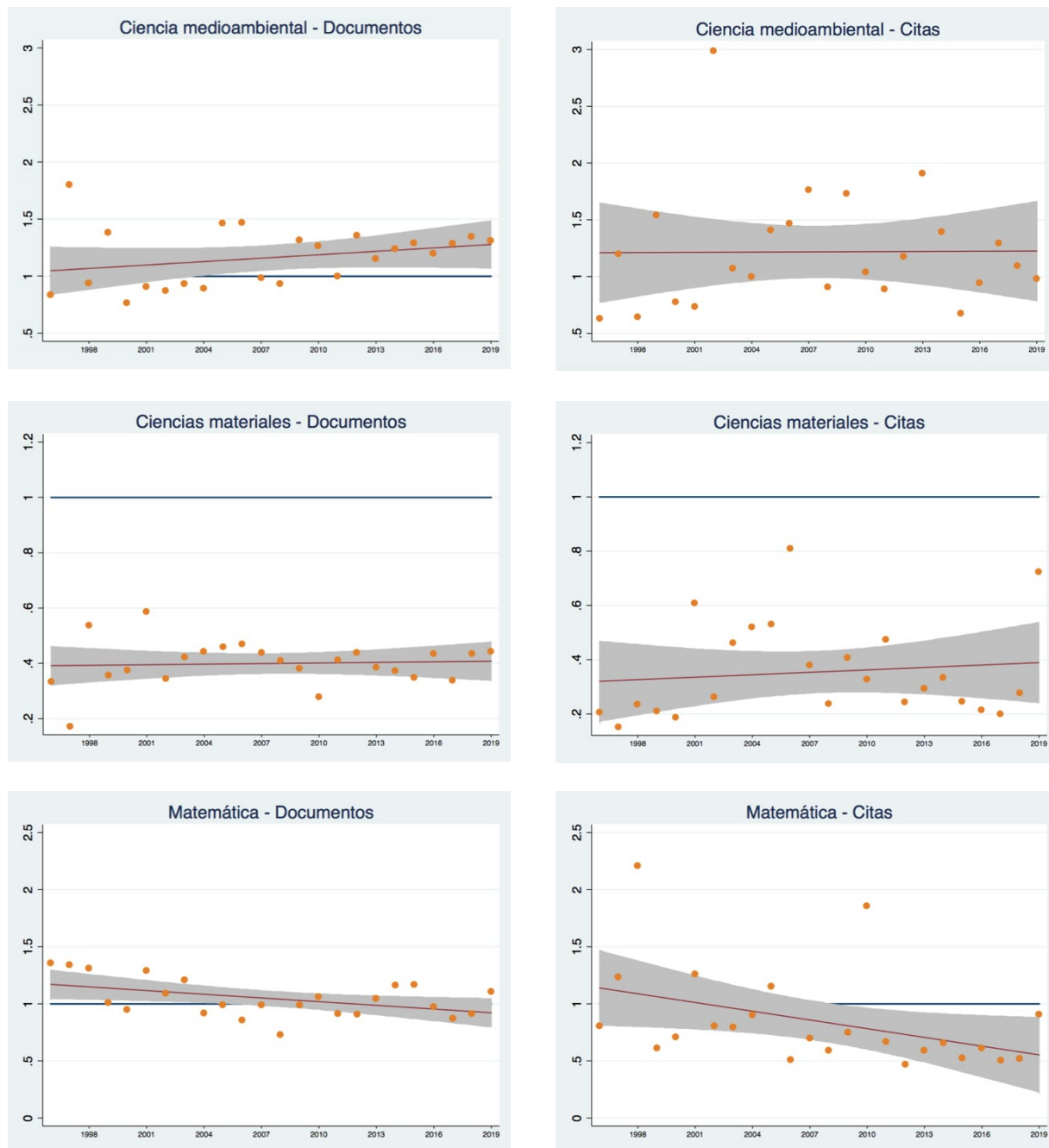

Nota: en gris intervalo de confianza al 95%

**Figura A6 (cont):** VCR por áreas temáticas con intervalo de confianza en Ciencias físicas

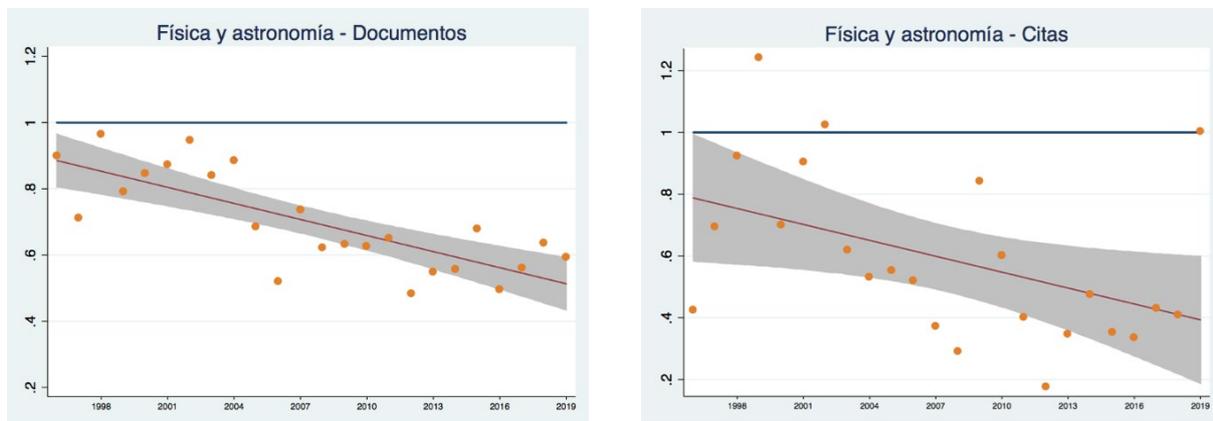

Nota: en gris intervalo de confianza al 95%

**Figura A7.** VCR por áreas temáticas con intervalo de confianza en Ciencias sociales

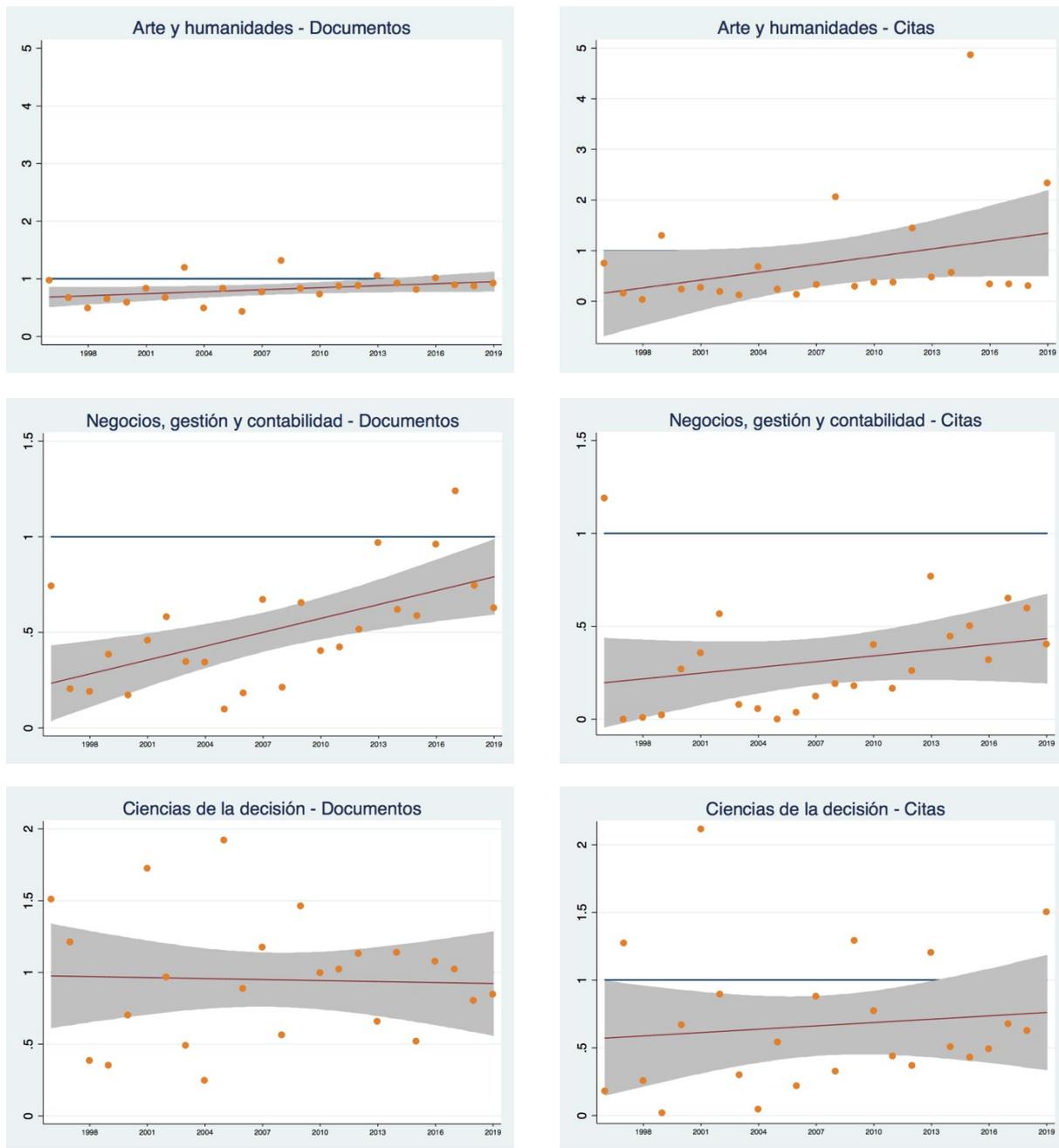

Nota: en gris intervalo de confianza al 95%

**Figura A7 (cont):** VCR por áreas temáticas con intervalo de confianza en Ciencias sociales

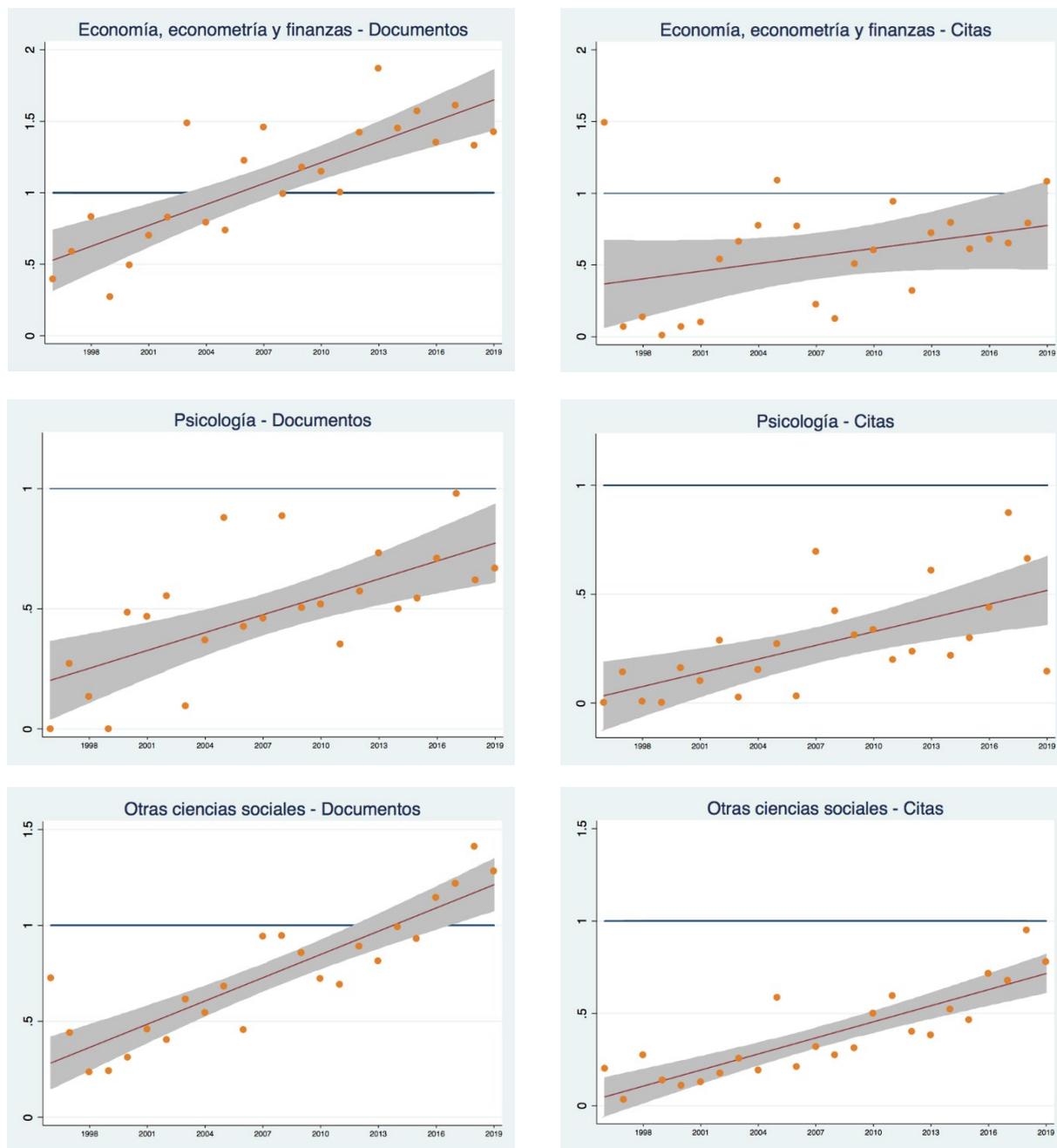

Nota: en gris intervalo de confianza al 95%

**Figura A8.** VCR por áreas temáticas con intervalo de confianza en Ciencias de la vida

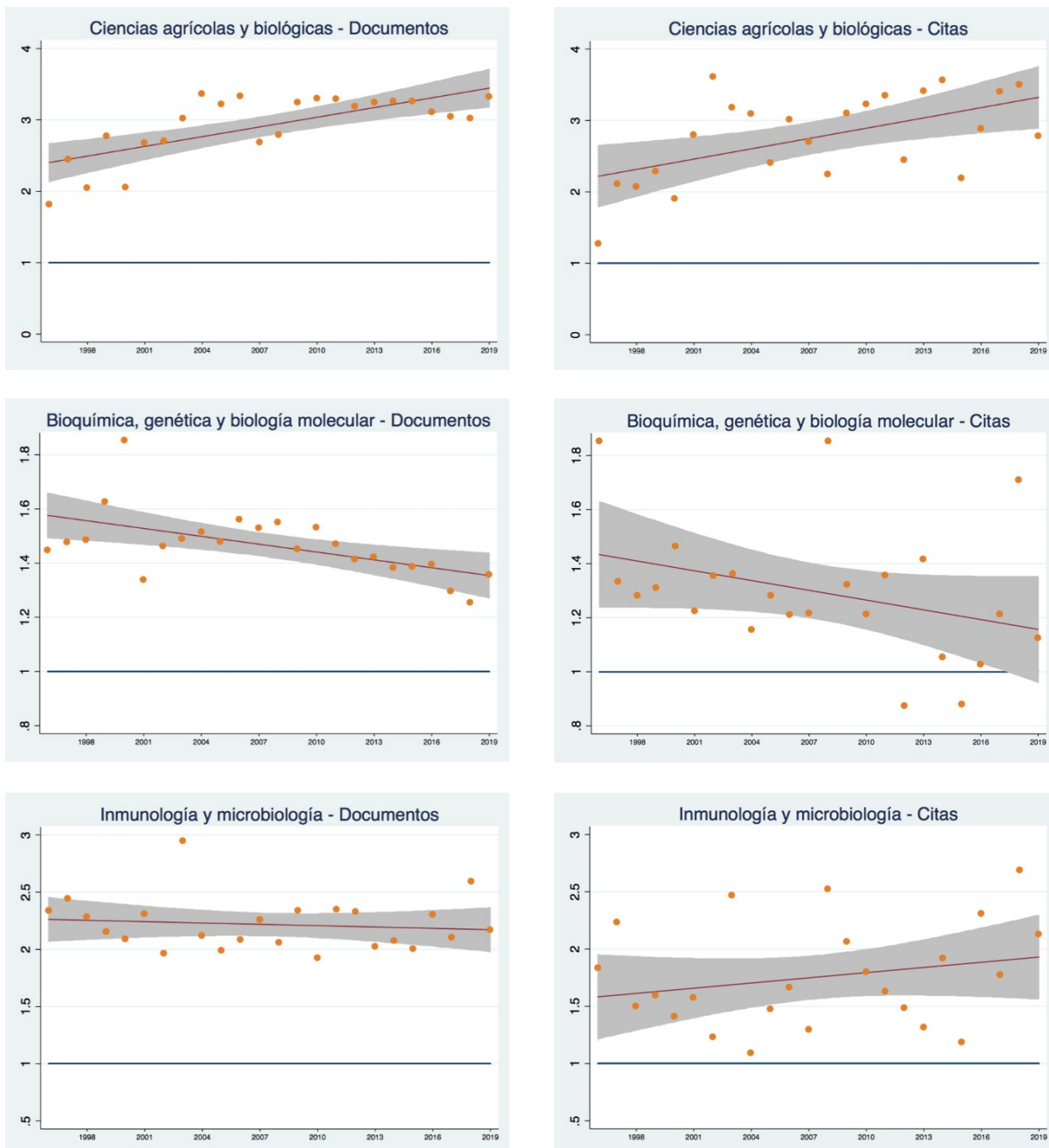

Nota: en gris intervalo de confianza al 95%

**Figura A8 (cont).** VCR por áreas temáticas con intervalo de confianza en Ciencias de la vida

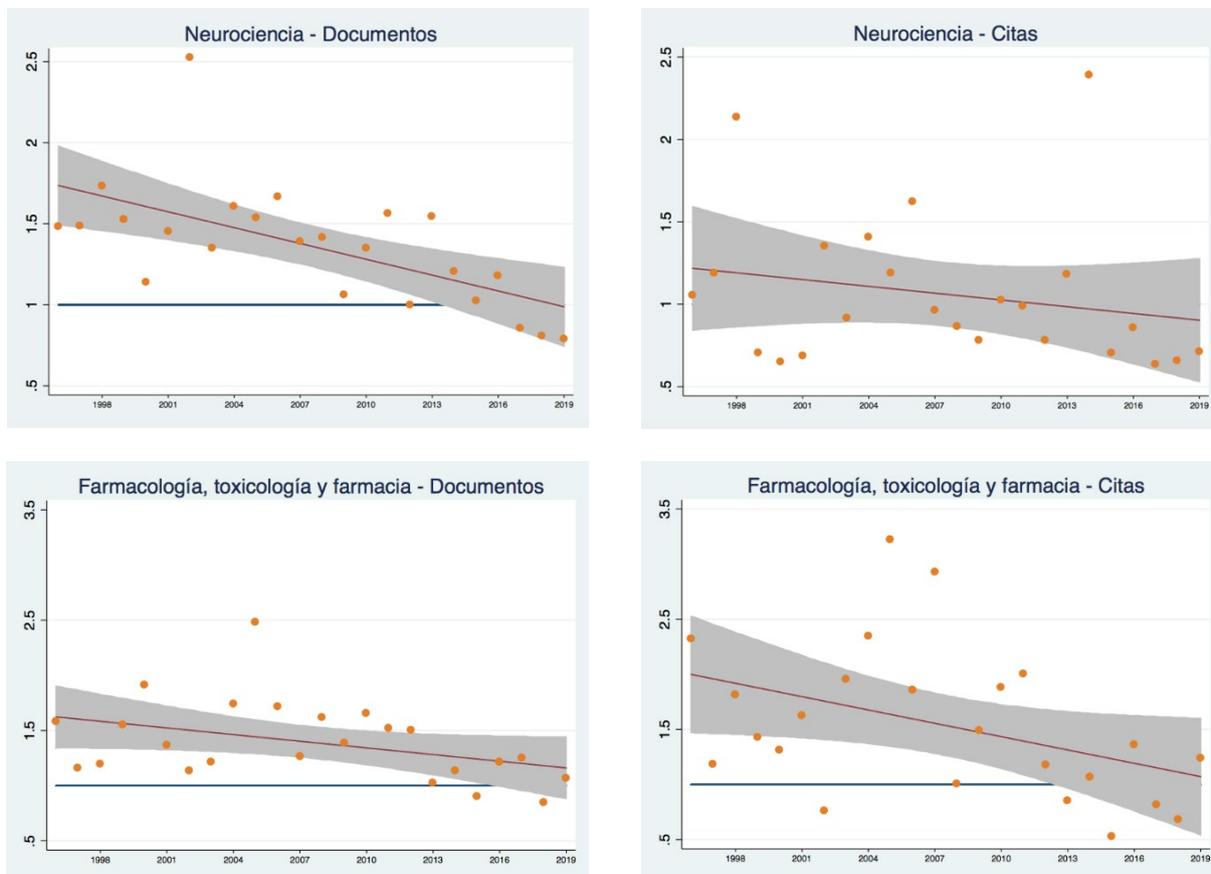

Nota: en gris intervalo de confianza al 95%

**Figura A9.** VCR por áreas temáticas con intervalo de confianza en Ciencias de la salud

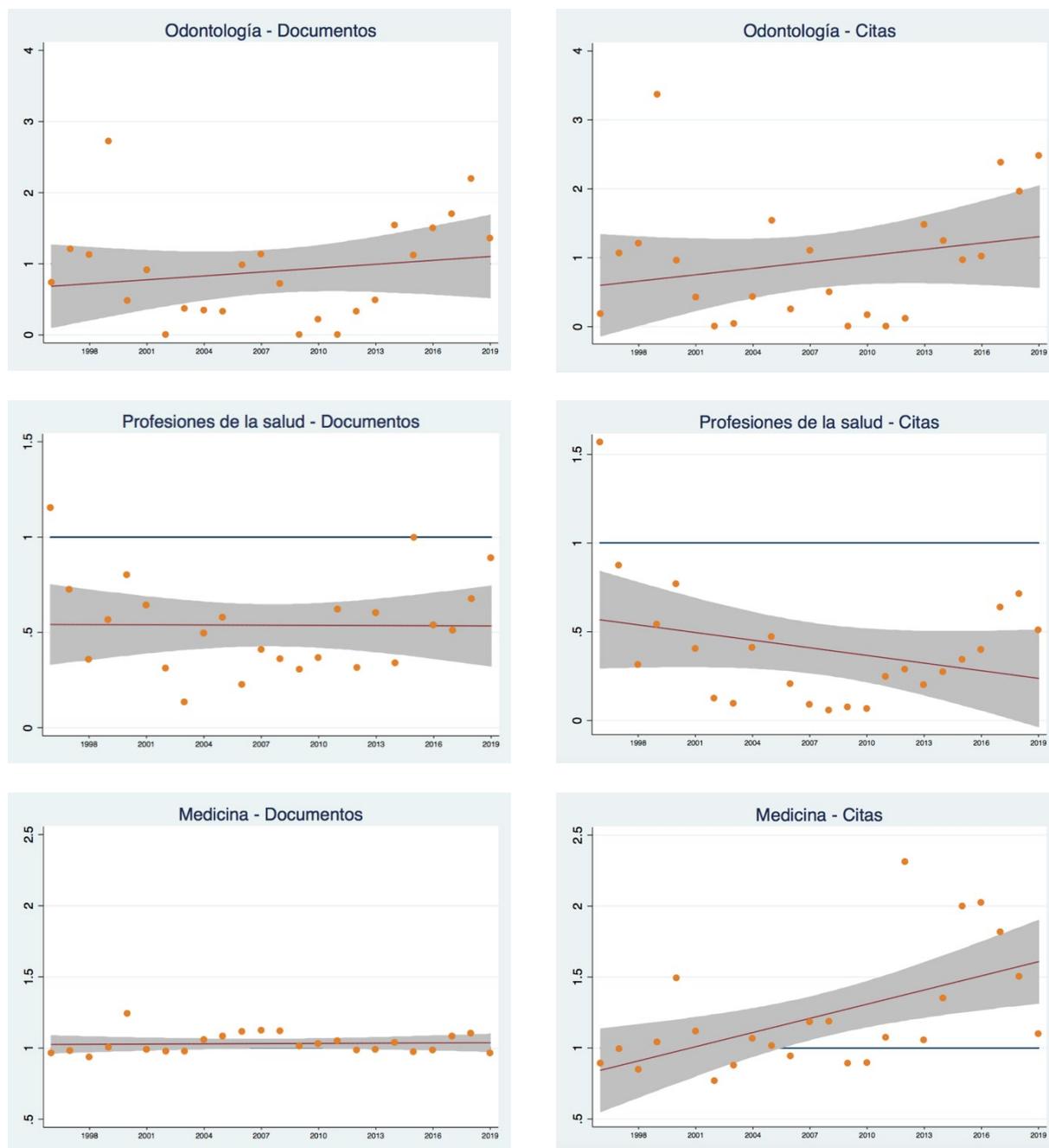

Nota: en gris intervalo de confianza al 95%

**Figura A9 (cont).** VCR por áreas temáticas con intervalo de confianza en Ciencias de la salud

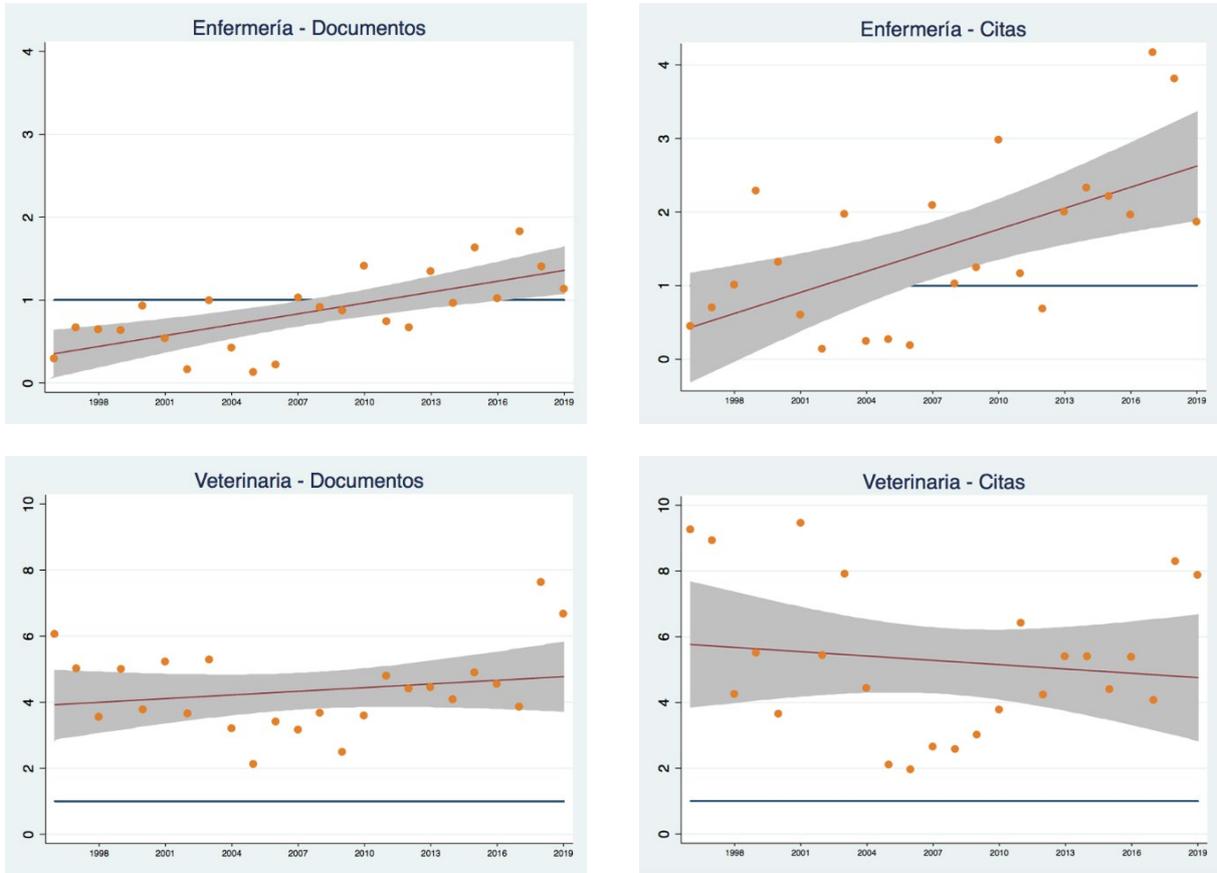

Nota: en gris intervalo de confianza al 95%

**Tabla A1:** VCR por grandes áreas (trienio 2017-2019)

|  | **Documentos** | | | **Citas** | | |
|---|---|---|---|---|---|---|
|  | Proporción | | VCR | Proporción | | VCR |
|  | Mundo | Uruguay |  | Mundo | Uruguay |  |
| Ciencias de la salud | 18,3% | 21,5% | 1,2 | 19,5% | 30,8% | 1,6 |
| Ciencias de la vida | 15,2% | 27,9% | 1,8 | 17,9% | 30,6% | 1,7 |
| Ciencias físicas | 54,1% | 37,1% | 0,7 | 53,5% | 30,6% | 0,6 |
| Ciencias sociales | 11,1% | 12,0% | 1,1 | 7,1% | 5,5% | 0,8 |

Nota: la suma de las proporciones da por debajo de 1 debido a que no se reportan las multidisciplinarias.

**Tabla A2.** Ventajas comparativas por disciplinas

| Grandes Áreas<br>Áreas<br>Disciplinas | Documentos | | | | Citas | | | |
|---|---|---|---|---|---|---|---|---|
| | % mundial | VCR trienio | VCR regresión | | % mundial | VCR trienio | VCR regresión | |
| Ciencias de la salud | 18% | 1,2 | 1,1 | *** | 19% | 1,6 | 1,6 | *** |
| Enfermería | 1% | 1,5 | 1,4 | *** | 1% | 3,2 | 2,6 | *** |
| Asistencia de enfermería | 0,002% | | | | 0,000% | | | |
| Atención comunitaria y domiciliaria | 0,02% | 0,4 | 0,7 | ** | 0,01% | 1,4 | 0,3 | *** |
| Enfermeras prácticas con licencia y enfermería vocacional con licencia | 0,01% | | 1,2 | | 0,01% | | 0,6 | |
| Enfermería (miscelánea) | 0,1% | 0,3 | 0,5 | *** | 0,1% | 0,3 | 0,3 | *** |
| Enfermería avanzada y especializada | 0,1% | | 0,8 | | 0,1% | | 0,2 | *** |
| Enfermería de cuidados intensivos | 0,01% | | 0,2 | ** | 0,004% | | 0,2 | *** |
| Enfermería de emergencia | 0,02% | | 0,1 | ** | 0,01% | | 0,1 | ** |
| Enfermería médica y quirúrgica | 0,01% | | 1,3 | | 0,004% | | 0,8 | |
| Evaluación y diagnóstico | 0,01% | | 2,1 | | 0,002% | | 3,7 | |
| Farmacología (enfermería) | 0,005% | | | | 0,000% | | | |
| Fundamentos y habilidades | 0,01% | 0,6 | 0,4 | *** | 0,004% | | 0,3 | |
| Gerontología | 0,04% | 0,4 | 0,4 | *** | 0,03% | 1,0 | 0,6 | |
| Investigación y teoría | 0,01% | 1,2 | 1,4 | * | 0,002% | | | |
| Liderazgo y gestión | 0,02% | | 0,6 | ** | 0,01% | | 0,5 | ** |
| Maternidad y partería | 0,02% | | 0,4 | *** | 0,01% | | 0,4 | *** |
| Nutrición y dietética | 0,2% | 3,8 | 3,2 | *** | 0,2% | 6,2 | 5,0 | *** |
| Oncología (enfermería) | 0,02% | 0,5 | 1,2 | | 0,01% | | 0,0 | *** |
| Pediatría | 0,01% | | 1,2 | | 0,01% | | 0,3 | * |
| Planificación de cuidados | 0,004% | | | | 0,002% | | | |
| Problemas, ética y aspectos legales | 0,03% | 0,3 | 1,1 | | 0,02% | | 1,7 | * |
| Psiquiatría y salud mental | 0,5% | 0,5 | 0,4 | *** | 0,5% | 0,5 | 0,3 | *** |
| Revisión y preparación de exámenes | 0,004% | | | | 0,001% | | | |
| Medicina | 16% | 1,0 | 1,0 | | 18% | 1,5 | 1,6 | *** |
| Anatomía | 0,1% | 2,7 | 3,4 | *** | 0,05% | 1,5 | 1,5 | |
| Anestesiología y medicina del dolor | 0,2% | 0,8 | 0,7 | * | 0,2% | 1,0 | 0,6 | |
| Bioquímica (médica) | 0,1% | 1,3 | 0,3 | | 0,1% | 1,2 | 0,7 | |
| Cardiología y medicina cardiovascular | 0,7% | 0,8 | 0,8 | ** | 1,0% | 0,6 | 0,7 | ** |

| Grandes Áreas<br>Áreas<br>Disciplinas | Documentos | | | | Citas | | | |
|---|---|---|---|---|---|---|---|---|
| Cirugía | 0,8% | 0,8 | 0,8 | ** | 0,5% | 0,6 | 0,6 | *** |
| Dermatología | 0,2% | 1,0 | 1,4 | | 0,1% | 0,7 | 2,0 | *** |
| Embriología | 0,01% | | 0,4 | *** | 0,01% | | 0,4 | *** |
| Endocrinología, diabetes y metabolismo | 0,3% | 0,6 | 1,8 | *** | 0,4% | 0,4 | 1,2 | |
| Enfermedades infecciosas | 0,6% | 2,2 | 0,4 | *** | 0,6% | 1,7 | 0,4 | *** |
| Epidemiología | 0,2% | 1,0 | 0,5 | | 0,2% | 1,0 | 0,7 | |
| Farmacología (médica) | 0,4% | 0,6 | 0,6 | *** | 0,4% | 1,3 | 0,9 | |
| Fisiología (médica) | 0,2% | 1,1 | 1,5 | *** | 0,3% | 1,6 | 2,3 | *** |
| Gastroenterología | 0,2% | 1,1 | 1,1 | * | 0,3% | 0,9 | | |
| Genética (clínica) | 0,2% | 1,3 | 2,4 | *** | 0,2% | 1,5 | 1,9 | *** |
| Geriatría y gerontología | 0,1% | 0,5 | 0,5 | *** | 0,2% | 0,3 | 0,1 | *** |
| Guía de medicamentos | 0,001% | | 1,9 | ** | 0,000% | | 1,9 | |
| Hematología | 0,3% | 1,0 | 0,9 | | 0,4% | 2,1 | 0,6 | |
| Hepatología | 0,1% | 1,2 | 1,0 | | 0,3% | 2,0 | 1,7 | |
| Histología | 0,1% | 2,1 | 0,8 | | 0,1% | 16,0 | 1,2 | |
| Informática de la salud | 0,2% | 0,8 | 0,9 | | 0,1% | 0,1 | 0,7 | |
| Inmunología y alergia | 0,4% | 1,2 | 1,2 | | 0,6% | 1,2 | 1,5 | |
| Medicina (miscelánea) | 4,8% | 1,0 | | | 5,8% | 1,5 | 0,0 | *** |
| Medicina alternativa y complementaria | 0,1% | 0,1 | 1,4 | | 0,1% | | 2,1 | |
| Medicina de cuidados intensivos y cuidados intensivos | 0,1% | 3,3 | | | 0,2% | 9,1 | | |
| Medicina de emergencia | 0,1% | 0,5 | | | 0,1% | 0,8 | | |
| Medicina interna | 0,2% | 1,2 | 0,6 | *** | 0,3% | 1,0 | 0,7 | |
| Medicina pulmonar y respiratoria | 0,3% | 2,1 | 2,0 | *** | 0,3% | 8,9 | 4,6 | ** |
| Medicina reproductiva | 0,1% | 1,3 | 1,2 | | 0,1% | 1,6 | 1,4 | |
| Microbiología (médica) | 0,3% | 2,1 | 3,1 | *** | 0,3% | 2,0 | 2,9 | *** |
| Nefrología | 0,1% | 2,6 | 2,4 | *** | 0,1% | 6,5 | 3,7 | *** |
| Neurología clínica | 0,6% | 0,5 | 0,5 | *** | 0,7% | 0,4 | 1,5 | |
| Obstetricia y ginecología | 0,3% | 1,4 | 1,4 | * | 0,2% | 0,9 | 1,0 | |
| Oftalmología | 0,2% | 0,1 | 0,6 | *** | 0,1% | 0,2 | 0,2 | *** |
| Oncología | 0,8% | 0,6 | 0,3 | *** | 1,2% | 0,7 | 0,1 | *** |
| Ortopedia y medicina deportiva | 0,4% | 0,6 | 0,5 | *** | 0,3% | 0,8 | 0,4 | *** |
| Otorrinolaringología | 0,2% | 1,1 | 1,0 | | 0,1% | 1,6 | 1,5 | |
| Patología y medicina forense | 0,2% | 1,0 | 0,7 | | 0,2% | 0,7 | 0,6 | ** |

| Grandes Áreas / Áreas / Disciplinas | Documentos | | | | Citas | | | |
|---|---|---|---|---|---|---|---|---|
| Pediatría, perinatología y salud infantil | 0,4% | 0,7 | 0,6 | * | 0,3% | 0,6 | 0,8 | |
| Política de salud | 0,2% | 0,9 | 2,1 | * | 0,2% | 1,1 | 0,4 | |
| Práctica familiar | 0,04% | 0,4 | 0,6 | *** | 0,02% | 0,8 | 0,4 | *** |
| Radiología, medicina nuclear e imágenes | 0,6% | 0,8 | 0,6 | *** | 0,5% | 0,6 | 0,6 | ** |
| Rehabilitación | 0,1% | 0,3 | 0,3 | *** | 0,1% | 0,3 | 0,2 | *** |
| Reseñas y referencias (médicas) | 0,000% | | | | 0,000% | | | |
| Reumatología | 0,1% | 1,4 | 1,0 | | 0,2% | 0,9 | 1,1 | |
| Salud mental psiquiátrica | 0,03% | | | | 0,03% | | | |
| Salud pública, ambiental y salud ocupacional | 0,7% | 1,2 | 1,4 | ** | 0,5% | 1,3 | 1,7 | ** |
| Trasplantes | 0,1% | 1,9 | 1,9 | ** | 0,1% | 2,1 | 1,4 | |
| Urología | 0,2% | 0,5 | 0,4 | *** | 0,2% | 0,2 | 0,2 | *** |
| Odontología | 0,3% | 1,8 | 1,1 | | 0,2% | 2,3 | 1,3 | |
| Asistencia dental | | | 0,7 | | | | 0,4 | |
| Cirugía oral | 0,1% | 1,6 | 0,9 | | 0,03% | 1,7 | 0,7 | |
| Higiene dental | 0,001% | | | | 0,000% | | | |
| Odontología (miscelánea) | 0,1% | 2,1 | | | 0,1% | 3,0 | | |
| Ortodoncia | 0,02% | | | | 0,01% | | | |
| Periodoncia | 0,02% | 1,1 | 1,3 | *** | 0,02% | 1,0 | 0,7 | |
| Profesiones de la salud | 1% | 0,7 | 0,5 | | 1% | 0,6 | 0,2 | *** |
| Asistencia médica y transcripción | 0,001% | | 0,7 | | 0,000% | | 0,9 | |
| Cuidado respiratorio | 0,000% | | | | 0,000% | | | |
| Farmacia | 0,03% | | 1,1 | | 0,01% | | 0,6 | |
| Fisioterapia, terapia deportiva y rehabilitación | 0,2% | 0,4 | 0,3 | *** | 0,2% | 0,4 | 0,2 | |
| Gestión de la información sanitaria | 0,05% | 1,8 | 1,1 | | 0,04% | 0,5 | 0,5 | *** |
| Habla y audición | 0,05% | 0,2 | 0,8 | | 0,03% | | 0,1 | *** |
| Optometría | 0,01% | | | | 0,004% | | | |
| Podología | 0,003% | | | | 0,001% | | | |
| Profesiones de la salud (miscelánea) | 0,02% | 1,2 | 0,7 | *** | 0,01% | 4,2 | 1,4 | |
| Química física y teórica | 0,8% | 1,1 | 1,0 | | 0,9% | 1,4 | 1,2 | |
| Quiropráctica | 0,004% | | | | 0,001% | | | |
| Servicios médicos de emergencia | 0,002% | | 1,0 | | 0,000% | | 0,0 | |

| Grandes Áreas / Áreas / Disciplinas | Documentos | | | | Citas | | | |
|---|---|---|---|---|---|---|---|---|
| Tecnología de laboratorio médico | 0,03% | 1,5 | 1,1 | | 0,02% | 0,7 | 0,2 | *** |
| Tecnología radiológica y de ultrasonido | 0,1% | 1,2 | 0,9 | | 0,1% | 1,0 | 0,3 | ** |
| Terapia manual y complementaria | 0,01% | | 0,5 | ** | 0,003% | | -0,2 | * |
| Terapia ocupacional | 0,01% | | | | 0,003% | | | |
| Terminología médica | 0,001% | | | | 0,000% | | -0,2 | *** |
| Veterinaria | 0,4% | 6,0 | 4,8 | *** | 0,2% | 6,7 | 4,8 | *** |
| Animales de comida | 0,04% | 12,4 | 0,4 | *** | 0,02% | 14,7 | 0,2 | *** |
| Animales pequeños | 0,02% | 11,5 | 6,5 | *** | 0,01% | 18,0 | 8,3 | * |
| Equino | 0,01% | 12,0 | 1,1 | | 0,01% | 15,9 | 1,9 | |
| Veterinaria (miscelánea) | 0,2% | 5,3 | 4,4 | *** | 0,1% | 5,8 | 4,6 | *** |
| Ciencias de la vida | 15% | 1,8 | 2,0 | *** | 18% | 1,7 | 1,6 | *** |
| Bioquímica, genética y biología molecular | 6% | 1,3 | 1,4 | *** | 9% | 1,3 | 1,2 | |
| Biofísica | 0,3% | 1,3 | 0,8 | | 0,3% | 1,5 | 0,5 | |
| Biología celular | 0,6% | 1,3 | 1,3 | ** | 0,9% | 2,9 | 1,9 | ** |
| Biología estructural | 0,1% | 1,6 | 1,6 | ** | 0,1% | 1,0 | 1,0 | |
| Biología molecular | 1,0% | 1,5 | 0,5 | *** | 1,2% | 1,3 | 0,3 | *** |
| Bioquímica | 1,1% | 1,6 | 1,8 | *** | 1,5% | 1,9 | 1,8 | *** |
| Bioquímica clínica | 0,2% | 1,5 | 1,7 | *** | 0,2% | 2,8 | 2,4 | *** |
| Bioquímica, genética y biología molecular (miscelánea) | 0,6% | 0,8 | 1,6 | *** | 1,0% | 0,8 | 1,3 | * |
| Biotecnología | 0,5% | 1,3 | 1,4 | ** | 0,6% | 0,9 | 0,9 | |
| Endocrinología | 0,2% | 1,8 | 2,4 | *** | 0,3% | 1,6 | 3,4 | *** |
| Envejecimiento | 0,1% | 1,6 | 1,2 | | 0,1% | 0,9 | 0,7 | |
| Fisiología | 0,4% | 1,3 | 1,3 | | 0,4% | 1,2 | 0,9 | |
| Genética | 0,7% | 2,0 | 1,7 | *** | 1,0% | 2,1 | 1,1 | |
| Investigación sobre el cancer | 0,5% | 0,7 | 0,6 | *** | 0,8% | 0,5 | 0,1 | *** |
| Medicina molecular | 0,4% | 1,3 | 1,6 | *** | 0,4% | 1,0 | 1,1 | |
| Neurociencia del desarrollo | 0,04% | 0,2 | 1,5 | | 0,1% | 0,4 | 0,9 | |
| Ciencias agrícolas y biológicas | 4% | 3,1 | 3,4 | *** | 4% | 3,2 | 3,3 | *** |
| Agronomía y ciencia de cultivos | 0,4% | 3,4 | 3,3 | *** | 0,3% | 3,3 | 3,2 | *** |
| Ciencia acuática | 0,4% | 3,2 | 4,4 | *** | 0,3% | 2,7 | 4,2 | *** |
| Ciencia de alimentos | 0,5% | 3,6 | 7,8 | *** | 0,5% | 4,9 | 7,9 | ** |
| Ciencia de las plantas | 0,5% | 2,3 | 2,4 | *** | 0,5% | 2,6 | 2,3 | *** |
| Ciencia de los insectos | 0,1% | 4,9 | 1,7 | *** | 0,1% | 6,4 | 1,9 | *** |

| Grandes Áreas<br>Áreas<br>Disciplinas | Documentos | | | | Citas | | | |
|---|---|---|---|---|---|---|---|---|
| Ciencia del suelo | 0,2% | 1,6 | 1,5 | *** | 0,2% | 1,2 | 1,5 | |
| Ciencias agrícolas y biológicas (miscelánea) | 0,4% | 1,5 | 1,4 | | 0,3% | 1,5 | 2,1 | ** |
| Ecología, evolución, comportamiento y sistemática | 0,8% | 3,1 | 2,8 | *** | 0,8% | 3,1 | 2,8 | *** |
| Horticultura | 0,1% | 4,0 | 0,9 | | 0,1% | 5,3 | 2,7 | * |
| Silvicultura | 0,2% | 1,6 | 3,9 | *** | 0,2% | 1,9 | 5,6 | *** |
| Zoología y ciencia animal | 0,4% | 5,8 | 6,5 | *** | 0,2% | 6,2 | 6,8 | *** |
| Farmacología, toxicología y farmacia | 2% | 1,1 | 1,2 | | 2% | 0,9 | 1,1 | |
| Ciencia farmacéutica | 0,3% | 0,8 | 0,9 | | 0,3% | 0,6 | 0,9 | |
| Descubrimiento de medicamentos | 0,4% | 1,3 | 1,5 | | 0,4% | 0,9 | 1,4 | |
| Farmacología | 0,5% | 1,0 | 1,1 | | 0,6% | 0,9 | 1,1 | |
| Farmacología, toxicología y farmacia (miscelánea) | 0,2% | 0,8 | 0,8 | | 0,1% | 1,2 | 2,8 | |
| Toxicología | 0,2% | 1,6 | 0,9 | | 0,2% | 1,5 | 0,3 | |
| Inmunología y microbiología | 2% | 2,3 | 2,2 | | 2% | 2,2 | 1,9 | *** |
| Inmunología | 0,4% | 1,6 | 0,7 | | 0,8% | 2,1 | 0,3 | *** |
| Inmunología y microbiología (miscelánea) | 0,2% | 1,4 | 1,3 | *** | 0,2% | 1,1 | 1,0 | |
| Microbiología | 0,3% | 3,3 | 0,2 | *** | 0,4% | 3,2 | 0,3 | *** |
| Microbiología y biotecnología aplicadas | 0,2% | 2,1 | 1,8 | ** | 0,2% | 2,2 | 1,6 | |
| Parasitología | 0,1% | 5,2 | 4,2 | *** | 0,1% | 5,1 | 3,7 | *** |
| Virología | 0,1% | 2,6 | 2,9 | *** | 0,2% | 1,7 | 2,5 | ** |
| Neurociencia | 2% | 0,8 | 1,0 | | 2% | 0,7 | 0,9 | |
| Neurociencia (miscelánea) | 0,3% | 0,7 | 0,9 | | 0,4% | 0,6 | 0,6 | * |
| Neurociencia celular y molecular | 0,2% | 1,0 | 1,2 | | 0,3% | 0,6 | 0,8 | |
| Neurociencia cognitiva | 0,2% | 0,6 | 0,6 | ** | 0,2% | 0,4 | 0,4 | *** |
| Neurociencia del comportamiento | 0,1% | 1,8 | 2,2 | *** | 0,1% | 1,8 | 1,7 | * |
| Neurología | 0,3% | 0,6 | 1,5 | | 0,4% | 0,6 | 3,0 | |
| Psicología del desarrollo y la educación | 0,2% | 0,3 | 1,5 | | 0,2% | 0,2 | 0,8 | |
| Psiquiatría biológica | 0,1% | 0,4 | 0,6 | | 0,1% | 0,3 | 0,2 | *** |
| Sistemas endócrinos y autónomos | 0,03% | 1,8 | 1,1 | | 0,04% | 5,2 | | |
| Sistemas sensoriales | 0,1% | 1,1 | 1,5 | ** | 0,1% | 1,1 | 1,7 | * |
| Farmacología, toxicología y farmacia | 2% | 1,1 | 1,2 | | 2% | 0,9 | 1,1 | |

| Grandes Áreas<br>Áreas<br>Disciplinas | Documentos | | | | Citas | | | |
|---|---|---|---|---|---|---|---|---|
| Ciencia farmacéutica | 0,3% | 0,8 | 0,9 | | 0,3% | 0,6 | 0,9 | |
| Descubrimiento de medicamentos | 0,4% | 1,3 | 1,5 | | 0,4% | 0,9 | 1,4 | |
| Farmacología | 0,5% | 1,0 | 1,1 | | 0,6% | 0,9 | 1,1 | |
| Farmacología, toxicología y farmacia (miscelánea) | 0,2% | 0,8 | 0,8 | | 0,1% | 1,2 | 2,8 | |
| Toxicología | 0,2% | 1,6 | 0,9 | | 0,2% | 1,5 | 0,3 | |
| Inmunología y microbiología | 2% | 2,3 | 2,2 | | 2% | 2,2 | 1,9 | *** |
| Inmunología | 0,4% | 1,6 | 0,7 | | 0,8% | 2,1 | 0,3 | *** |
| Inmunología y microbiología (miscelánea) | 0,2% | 1,4 | 1,3 | *** | 0,2% | 1,1 | 1,0 | |
| Microbiología | 0,3% | 3,3 | 0,2 | *** | 0,4% | 3,2 | 0,3 | *** |
| Microbiología y biotecnología aplicadas | 0,2% | 2,1 | 1,8 | ** | 0,2% | 2,2 | 1,6 | |
| Parasitología | 0,1% | 5,2 | 4,2 | *** | 0,1% | 5,1 | 3,7 | *** |
| Virología | 0,1% | 2,6 | 2,9 | *** | 0,2% | 1,7 | 2,5 | ** |
| Neurociencia | 2% | 0,8 | 1,0 | | 2% | 0,7 | 0,9 | |
| Neurociencia (miscelánea) | 0,3% | 0,7 | 0,9 | | 0,4% | 0,6 | 0,6 | * |
| Neurociencia celular y molecular | 0,2% | 1,0 | 1,2 | | 0,3% | 0,6 | 0,8 | |
| Neurociencia cognitiva | 0,2% | 0,6 | 0,6 | ** | 0,2% | 0,4 | 0,4 | *** |
| Neurociencia del comportamiento | 0,1% | 1,8 | 2,2 | *** | 0,1% | 1,8 | 1,7 | * |
| Neurología | 0,3% | 0,6 | 1,5 | | 0,4% | 0,6 | 3,0 | |
| Psicología del desarrollo y la educación | 0,2% | 0,3 | 1,5 | | 0,2% | 0,2 | 0,8 | |
| Psiquiatría biológica | 0,1% | 0,4 | 0,6 | | 0,1% | 0,3 | 0,2 | *** |
| Sistemas endócrinos y autónomos | 0,03% | 1,8 | 1,1 | | 0,04% | 5,2 | | |
| Sistemas sensoriales | 0,1% | 1,1 | 1,5 | ** | 0,1% | 1,1 | 1,7 | * |
| Farmacología, toxicología y farmacia | 2% | 1,1 | 1,2 | | 2% | 0,9 | 1,1 | |
| Ciencia farmacéutica | 0,3% | 0,8 | 0,9 | | 0,3% | 0,6 | 0,9 | |
| Descubrimiento de medicamentos | 0,4% | 1,3 | 1,5 | | 0,4% | 0,9 | 1,4 | |
| Farmacología | 0,5% | 1,0 | 1,1 | | 0,6% | 0,9 | 1,1 | |
| Farmacología, toxicología y farmacia (miscelánea) | 0,2% | 0,8 | 0,8 | | 0,1% | 1,2 | 2,8 | |
| Toxicología | 0,2% | 1,6 | 0,9 | | 0,2% | 1,5 | 0,3 | |
| Inmunología y microbiología | 2% | 2,3 | 2,2 | | 2% | 2,2 | 1,9 | *** |
| Inmunología | 0,4% | 1,6 | 0,7 | | 0,8% | 2,1 | 0,3 | *** |

| Grandes Áreas / Áreas / Disciplinas | Documentos | | | | Citas | | | |
|---|---|---|---|---|---|---|---|---|
| Inmunología y microbiología (miscelánea) | 0,2% | 1,4 | 1,3 | *** | 0,2% | 1,1 | 1,0 | |
| Microbiología | 0,3% | 3,3 | 0,2 | *** | 0,4% | 3,2 | 0,3 | *** |
| Microbiología y biotecnología aplicadas | 0,2% | 2,1 | 1,8 | ** | 0,2% | 2,2 | 1,6 | |
| Parasitología | 0,1% | 5,2 | 4,2 | *** | 0,1% | 5,1 | 3,7 | *** |
| Virología | 0,1% | 2,6 | 2,9 | *** | 0,2% | 1,7 | 2,5 | ** |
| Neurociencia | 2% | 0,8 | 1,0 | | 2% | 0,7 | 0,9 | |
| Neurociencia (miscelánea) | 0,3% | 0,7 | 0,9 | | 0,4% | 0,6 | 0,6 | * |
| Neurociencia celular y molecular | 0,2% | 1,0 | 1,2 | | 0,3% | 0,6 | 0,8 | |
| Neurociencia cognitiva | 0,2% | 0,6 | 0,6 | ** | 0,2% | 0,4 | 0,4 | *** |
| Neurociencia del comportamiento | 0,1% | 1,8 | 2,2 | *** | 0,1% | 1,8 | 1,7 | * |
| Neurología | 0,3% | 0,6 | 1,5 | | 0,4% | 0,6 | 3,0 | |
| Psicología del desarrollo y la educación | 0,2% | 0,3 | 1,5 | | 0,2% | 0,2 | 0,8 | |
| Psiquiatría biológica | 0,1% | 0,4 | 0,6 | | 0,1% | 0,3 | 0,2 | *** |
| Sistemas endócrinos y autónomos | 0,03% | 1,8 | 1,1 | | 0,04% | 5,2 | | |
| Sistemas sensoriales | 0,1% | 1,1 | 1,5 | ** | 0,1% | 1,1 | 1,7 | * |
| Farmacología, toxicología y farmacia | 2% | 1,1 | 1,2 | | 2% | 0,9 | 1,1 | |
| Ciencia farmacéutica | 0,3% | 0,8 | 0,9 | | 0,3% | 0,6 | 0,9 | |
| Descubrimiento de medicamentos | 0,4% | 1,3 | 1,5 | | 0,4% | 0,9 | 1,4 | |
| Farmacología | 0,5% | 1,0 | 1,1 | | 0,6% | 0,9 | 1,1 | |
| Farmacología, toxicología y farmacia (miscelánea) | 0,2% | 0,8 | 0,8 | | 0,1% | 1,2 | 2,8 | |
| Toxicología | 0,2% | 1,6 | 0,9 | | 0,2% | 1,5 | 0,3 | |
| Inmunología y microbiología | 2% | 2,3 | 2,2 | | 2% | 2,2 | 1,9 | *** |
| Inmunología | 0,4% | 1,6 | 0,7 | | 0,8% | 2,1 | 0,3 | *** |
| Inmunología y microbiología (miscelánea) | 0,2% | 1,4 | 1,3 | *** | 0,2% | 1,1 | 1,0 | |
| Microbiología | 0,3% | 3,3 | 0,2 | *** | 0,4% | 3,2 | 0,3 | *** |
| Microbiología y biotecnología aplicadas | 0,2% | 2,1 | 1,8 | ** | 0,2% | 2,2 | 1,6 | |
| Parasitología | 0,1% | 5,2 | 4,2 | *** | 0,1% | 5,1 | 3,7 | *** |
| Virología | 0,1% | 2,6 | 2,9 | *** | 0,2% | 1,7 | 2,5 | ** |
| Neurociencia | 2% | 0,8 | 1,0 | | 2% | 0,7 | 0,9 | |
| Neurociencia (miscelánea) | 0,3% | 0,7 | 0,9 | | 0,4% | 0,6 | 0,6 | * |
| Neurociencia celular y molecular | 0,2% | 1,0 | 1,2 | | 0,3% | 0,6 | 0,8 | |
| Neurociencia cognitiva | 0,2% | 0,6 | 0,6 | ** | 0,2% | 0,4 | 0,4 | *** |
| Neurociencia del comportamiento | 0,1% | 1,8 | 2,2 | *** | 0,1% | 1,8 | 1,7 | * |

| Grandes Áreas Áreas Disciplinas | Documentos | | | | Citas | | | |
|---|---|---|---|---|---|---|---|---|
| Neurología | 0,3% | 0,6 | 1,5 | | 0,4% | 0,6 | 3,0 | |
| Psicología del desarrollo y la educación | 0,2% | 0,3 | 1,5 | | 0,2% | 0,2 | 0,8 | |
| Psiquiatría biológica | 0,1% | 0,4 | 0,6 | | 0,1% | 0,3 | 0,2 | *** |
| Sistemas endócrinos y autónomos | 0,03% | 1,8 | 1,1 | | 0,04% | 5,2 | | |
| Sistemas sensoriales | 0,1% | 1,1 | 1,5 | ** | 0,1% | 1,1 | 1,7 | * |
| Ciencias físicas | 54% | 0,7 | 0,7 | *** | 54% | 0,6 | 0,5 | *** |
| Ciencia medioambiental | 4% | 1,3 | 1,3 | *** | 4% | 1,1 | 1,2 | |
| Cambio planetario y global | 0,1% | 1,7 | 0,8 | | 0,2% | 1,5 | 1,0 | |
| Ciencia y tecnología del agua | 0,4% | 0,9 | 0,8 | | 0,4% | 0,9 | 1,2 | |
| Ciencias ambientales (miscelánea) | 0,7% | 0,6 | 0,7 | * | 0,7% | 0,7 | 0,9 | |
| Conservación de la naturaleza y el paisaje | 0,2% | 2,4 | 1,3 | | 0,2% | 3,3 | 0,1 | *** |
| Contaminación | 0,5% | 1,0 | 0,8 | * | 0,8% | 0,7 | 0,8 | |
| Ecología | 0,5% | 2,8 | 1,7 | ** | 0,5% | 2,7 | 1,2 | |
| Gestión y eliminación de residuos | 0,3% | 1,4 | 0,9 | | 0,5% | 0,8 | 0,9 | |
| Gestión, seguimiento, política y derecho | 0,4% | 1,1 | 1,1 | | 0,5% | 0,6 | 0,8 | |
| Ingeniería ambiental | 0,4% | 0,9 | 0,5 | | 0,6% | 0,7 | 0,8 | |
| Modelado ecológico | 0,1% | 1,8 | 1,1 | | 0,1% | 1,2 | 1,2 | |
| Química ambiental | 0,5% | 0,9 | 0,4 | *** | 0,9% | 0,6 | 0,3 | *** |
| Salud, toxicología y mutagénesis | 0,3% | 1,3 | 1,4 | ** | 0,4% | 0,8 | 1,8 | |
| Ciencias de la computación | 8% | 0,8 | 0,9 | ** | 5% | 0,5 | 0,5 | *** |
| Aplicaciones de ciencia de la computación | 1,8% | 0,6 | 1,1 | | 1,4% | 0,4 | 0,6 | |
| Ciencia de la computación (miscelánea) | 1,1% | 1,0 | 0,8 | | 0,6% | 0,7 | 0,6 | |
| Gráficos por computadora y diseño asistido por computadora | 0,2% | 0,7 | 0,7 | ** | 0,1% | 0,3 | 0,8 | |
| Hardware y arquitectura | 0,7% | 1,0 | 1,7 | *** | 0,4% | 0,8 | 1,9 | ** |
| Inteligencia artificial | 0,9% | 0,7 | 0,6 | *** | 0,6% | 0,3 | 0,7 | |
| Interacción persona-ordenador | 0,4% | 1,0 | 1,1 | | 0,2% | 0,4 | 0,4 | |
| Procesamiento de señales | 0,6% | 0,8 | 0,7 | ** | 0,4% | 0,7 | 0,5 | *** |
| Redes informáticas y comunicaciones | 1,6% | 0,8 | 0,7 | ** | 0,8% | 0,8 | 0,6 | |
| Sistemas de información | 0,8% | 0,6 | 2,3 | *** | 0,6% | 0,3 | 2,2 | *** |
| Software | 1,1% | 1,2 | 1,3 | *** | 1,0% | 0,5 | 0,8 | |

| Grandes Áreas<br>Áreas<br>Disciplinas | Documentos | | | | Citas | | | |
|---|---|---|---|---|---|---|---|---|
| Teoría y matemáticas computacionales | 0,2% | 0,8 | 0,5 | *** | 0,2% | 0,4 | 0,3 | *** |
| Visión por computadora y reconocimiento de patrones | 0,5% | 0,8 | 0,6 | ** | 0,3% | 0,7 | 0,4 | *** |
| Ciencias materiales | 6% | 0,4 | 0,4 | *** | 7% | 0,4 | 0,4 | *** |
| Biomateriales | 0,3% | 0,4 | 0,2 | *** | 0,4% | 0,4 | 0,3 | *** |
| Cerámicas y materiales compuestos | 0,3% | 0,2 | 0,3 | *** | 0,4% | 0,3 | 0,3 | *** |
| Ciencia de los materiales (miscelánea) | 1,9% | 0,4 | 0,5 | *** | 2,5% | 0,4 | 0,5 | *** |
| Materiales electrónicos, ópticos y magnéticos | 1,4% | 0,4 | 0,7 | | 1,3% | 0,3 | 0,5 | ** |
| Metales y aleaciones | 0,4% | 0,2 | 1,0 | | 0,4% | 0,1 | 2,1 | *** |
| Polímeros y plásticos | 0,3% | 0,1 | 0,5 | *** | 0,3% | 0,0 | 0,2 | *** |
| Química de materiales | 0,9% | 0,5 | 0,6 | *** | 1,1% | 0,4 | 0,1 | *** |
| Superficies, recubrimientos y películas | 0,5% | 0,6 | 0,6 | *** | 0,7% | 0,6 | 0,8 | |
| Ciencias planetarias y de la tierra | 3% | 0,8 | 0,8 | ** | 3% | 0,7 | 0,7 | ** |
| Ciencia atmosférica | 0,3% | 0,5 | 0,4 | ** | 0,3% | 0,5 | 0,7 | |
| Ciencia espacial y planetaria | 0,5% | 0,3 | 0,2 | *** | 0,9% | 0,4 | 0,3 | *** |
| Ciencias de la tierra y planetarias (miscelánea) | 0,6% | 0,5 | 2,8 | | 0,4% | 0,6 | 12,3 | |
| Computadoras en ciencias de la tierra | 0,1% | 0,4 | 0,9 | | 0,1% | 0,2 | 1,2 | |
| Estratigrafía | 0,04% | 1,4 | 1,6 | | 0,04% | 0,9 | 1,2 | |
| Geofísica | 0,3% | 0,2 | 1,3 | | 0,3% | 0,1 | 0,9 | |
| Geología | 0,3% | 1,3 | 1,4 | *** | 0,3% | 1,6 | 1,2 | |
| Geología económica | 0,05% | 0,4 | 3,6 | *** | 0,04% | 0,4 | 2,9 | *** |
| Geoquímica y petrología | 0,3% | 0,5 | 1,4 | | 0,3% | 0,7 | 1,2 | |
| Ingeniería geotécnica e ingeniería en geología | 0,3% | 0,0 | 0,3 | *** | 0,2% | | 0,5 | * |
| Oceanografía | 0,2% | 2,1 | 2,9 | *** | 0,2% | 1,5 | 2,7 | *** |
| Paleontología | 0,1% | 2,1 | 2,4 | ** | 0,1% | 0,9 | 1,5 | |
| Procesos de la superficie terrestre | 0,2% | 1,2 | 0,4 | *** | 0,2% | 1,2 | 0,6 | |
| Energía | 3% | 0,6 | 0,6 | *** | 3% | 0,4 | 0,4 | ** |
| Energía (miscelánea) | 0,4% | 0,5 | 0,4 | *** | 0,5% | 0,3 | 0,3 | *** |
| Energía renovable, sostenibilidad y medio ambiente | 0,8% | 0,8 | 0,8 | | 1,4% | 0,4 | 0,7 | |
| Física nuclear y de altas energías | 0,3% | 0,6 | 0,2 | *** | 0,5% | 0,7 | 0,3 | ** |

| Grandes Áreas Áreas Disciplinas | Documentos | | | | Citas | | | |
|---|---|---|---|---|---|---|---|---|
| Ingeniería energética y tecnología energética | 1,0% | 0,7 | 0,2 | *** | 0,9% | 0,5 | 0,6 | *** |
| Tecnología de combustible | 0,4% | 0,4 | 1,6 | ** | 0,6% | 0,4 | 1,9 | * |
| Física y astronomía | 7% | 0,6 | 0,5 | *** | 8% | 0,6 | 0,4 | *** |
| Acústica y ultrasonidos | 0,1% | 0,8 | 0,7 | | 0,1% | 0,9 | 1,5 | |
| Astronomía y astrofísica | 0,4% | 0,4 | 0,1 | *** | 0,8% | 0,5 | 0,1 | *** |
| Energía e ingeniería nuclear | 0,2% | 0,2 | 0,4 | *** | 0,3% | 0,1 | 0,1 | *** |
| Física atómica, molecular y óptica | 0,8% | 0,5 | 0,4 | *** | 0,8% | 0,5 | 0,1 | ** |
| Física de la materia condensada | 1,7% | 0,6 | 0,9 | | 1,8% | 0,5 | 0,1 | |
| Física estadística y no lineal | 0,1% | 1,4 | 1,2 | | 0,1% | 0,7 | 1,0 | |
| Física y astronomía (miscelánea) | 1,5% | 0,7 | 0,6 | *** | 2,1% | 1,0 | 0,5 | ** |
| Instrumentación | 0,8% | 0,6 | 4,8 | *** | 0,5% | 0,3 | 4,8 | *** |
| Radiación | 0,2% | 0,3 | 0,5 | *** | 0,1% | 0,2 | 0,3 | |
| Superficies e interfaces | 0,2% | 0,7 | 0,5 | *** | 0,2% | 0,9 | 0,7 | |
| Ingeniería | 11% | 0,4 | 0,5 | *** | 9% | 0,3 | 0,3 | *** |
| Arquitectura | 0,1% | 1,0 | 0,7 | | 0,03% | 0,3 | 0,3 | *** |
| Construcción y edificación | 0,4% | 0,6 | 0,4 | *** | 0,4% | 0,2 | 0,4 | * |
| Ingeniería (miscelánea) | 1,2% | 0,4 | 0,5 | *** | 0,9% | 0,6 | 0,4 | ** |
| Ingenieria aeroespacial | 0,3% | 0,0 | 0,4 | *** | 0,2% | 0,0 | 0,1 | *** |
| Ingeniería automotriz | 0,2% | 0,1 | 0,7 | ** | 0,1% | 0,1 | 0,0 | *** |
| Ingeniería biomédica | 0,5% | 0,8 | 1,0 | | 0,5% | 0,5 | 0,6 | *** |
| Ingeniería civil y estructural | 0,7% | 0,4 | 0,4 | *** | 0,7% | 0,2 | 0,3 | *** |
| Ingeniería eléctrica y electrónica | 2,7% | 0,5 | 1,2 | | 2,2% | 0,3 | 0,8 | ** |
| Ingeniería en sistemas y control | 0,9% | 0,1 | 0,7 | | 0,7% | 0,1 | 0,8 | |
| Ingeniería mecánica | 1,6% | 0,2 | 1,4 | | 1,6% | 0,1 | 1,0 | |
| Ingeniería oceánica | 0,2% | 0,3 | 0,5 | *** | 0,1% | 0,4 | 0,7 | |
| Mecánica computacional | 0,1% | 0,7 | 0,9 | | 0,1% | 0,2 | 0,4 | ** |
| Mecánica de materiales | 1,0% | 0,3 | 0,2 | *** | 1,1% | 0,1 | 0,1 | *** |
| Relaciones industriales | 0,03% | 2,0 | 1,4 | | 0,01% | 0,4 | 1,1 | |
| Seguridad, riesgo, confiabilidad y calidad | 0,5% | 0,6 | 0,5 | *** | 0,2% | 0,6 | 0,7 | |
| Tecnología de medios | 0,2% | 0,8 | 0,2 | *** | 0,1% | 0,7 | 0,1 | *** |
| Ingeniería química | 3% | 0,7 | 0,6 | *** | 4% | 0,4 | 0,3 | *** |
| Bioingeniería | 0,4% | 1,0 | 0,7 | | 0,5% | 0,6 | 0,2 | *** |
| Catálisis | 0,4% | 0,9 | 1,0 | | 0,9% | 0,4 | 0,4 | *** |

| Grandes Áreas<br>Áreas<br>Disciplinas | Documentos | | | | Citas | | | |
|---|---|---|---|---|---|---|---|---|
| Filtración y separación | 0,04% | | 0,8 | *** | 0,1% | | 0,6 | ** |
| Ingeniería química (miscelánea) | 0,9% | 0,6 | 0,4 | *** | 1,0% | 0,4 | 0,2 | *** |
| Procesos de flujo y transferencia de fluidos | 0,2% | 0,4 | 0,8 | * | 0,2% | 0,4 | 0,0 | ** |
| Química coloide y de superficie | 0,1% | 0,2 | 0,5 | *** | 0,3% | | 0,2 | *** |
| Química y tecnología de procesos | 0,2% | 0,7 | 0,5 | ** | 0,3% | 0,6 | 0,1 | * |
| Salud y seguridad química | 0,01% | | 1,3 | | 0,01% | | 0,4 | |
| Matemática | 4% | 1,0 | 0,9 | | 3% | 0,7 | 0,6 | *** |
| Álgebra y teoría de números | 0,1% | 2,1 | 2,1 | ** | 0,04% | 1,6 | 1,7 | * |
| Analisis | 0,2% | 0,8 | 0,9 | | 0,1% | 0,3 | 0,5 | |
| Análisis numérico | 0,1% | 0,7 | 1,0 | | 0,1% | 3,2 | 1,9 | |
| Control y optimización | 0,6% | 0,4 | 0,9 | | 0,2% | 0,8 | 0,9 | |
| Estadística y probabilidad | 0,3% | 1,5 | 0,8 | | 0,2% | 1,6 | 1,0 | |
| Física matemática | 0,1% | 1,3 | 1,0 | | 0,1% | 0,6 | 0,3 | *** |
| Geometría y topología | 0,1% | 1,5 | 1,6 | ** | 0,04% | 1,8 | 1,5 | |
| Informática teórica | 0,5% | 1,0 | 1,4 | ** | 0,3% | 0,9 | 0,7 | |
| Lógica | 0,05% | 0,7 | 0,5 | | 0,02% | 0,9 | 0,6 | |
| Matemáticas (miscelánea) | 0,5% | 1,8 | 1,1 | | 0,2% | 1,3 | 0,5 | |
| Matemáticas aplicadas | 0,9% | 0,8 | 0,8 | ** | 0,6% | 0,3 | 0,5 | |
| Matemáticas computacionales | 0,2% | 1,3 | 1,4 | | 0,2% | 0,2 | 1,1 | |
| Matemáticas discretas y combinatorias | 0,1% | 1,4 | 0,2 | *** | 0,03% | 1,2 | 0,4 | *** |
| Modelado y simulación | 0,6% | 0,6 | 2,0 | *** | 0,4% | 0,3 | 1,8 | ** |
| Química | 5% | 0,8 | 0,9 | * | 7% | 0,8 | 0,8 | ** |
| Electroquímica | 0,2% | 0,6 | 0,6 | *** | 0,3% | 0,3 | 0,2 | *** |
| Espectroscopía | 0,3% | 0,9 | 1,0 | | 0,3% | 0,8 | 1,2 | |
| Química (miscelánea) | 1,7% | 0,6 | 0,5 | *** | 2,9% | 0,6 | 0,5 | *** |
| Química analítica | 0,4% | 1,3 | 1,3 | | 0,5% | 1,1 | 1,4 | |
| Química inorgánica | 0,3% | 1,4 | 0,5 | ** | 0,4% | 1,2 | 0,4 | *** |
| Química orgánica | 0,7% | 1,2 | 1,4 | ** | 0,8% | 1,2 | 1,2 | |
| Ciencias sociales | 11% | 1,1 | 1,1 | *** | 7% | 0,8 | 0,8 | *** |
| Arte y humanidades | 2% | 0,9 | 1,0 | * | 1% | 1,0 | 1,3 | |
| Arqueología (arte y humanidades) | 0,1% | 1,4 | 1,0 | | 0,05% | 1,2 | 1,1 | |
| Arte y humanidades (miscelánea) | 0,3% | 0,9 | 0,9 | | 0,2% | 0,3 | 0,5 | *** |
| Artes visuales y escénicas | 0,1% | 0,6 | 0,8 | | 0,01% | 2,0 | 0,8 | |

| Grandes Áreas<br>Áreas<br>Disciplinas | Documentos | | | | Citas | | | |
|---|---|---|---|---|---|---|---|---|
| Clásicos | 0,02% | | | | 0,001% | | | |
| Conservación | 0,03% | 0,6 | 0,5 | ** | 0,01% | 0,6 | 0,5 | *** |
| Estudios religiosos | 0,1% | 0,4 | 0,5 | *** | 0,01% | 1,2 | 0,7 | |
| Filosofía | 0,2% | 1,0 | 1,0 | | 0,05% | 0,5 | 0,6 | |
| Historia | 0,3% | 1,1 | 1,7 | | 0,05% | 0,9 | 6,8 | ** |
| Historia y filosofía de la ciencia | 0,1% | 1,0 | 1,3 | * | 0,3% | 1,9 | 1,4 | |
| Lengua y lingüística | 0,2% | 0,7 | | | 0,1% | 0,0 | | |
| Literatura y teoría literaria | 0,2% | 0,7 | 0,4 | ** | 0,01% | | 0,1 | *** |
| Museología | 0,01% | 1,2 | 0,9 | | 0,002% | 2,4 | 1,3 | |
| Música | 0,03% | 0,7 | 1,7 | | 0,01% | | 2,7 | * |
| Ciencias de la decisión | 1% | 0,9 | 0,9 | | 1% | 1,0 | 0,8 | |
| Ciencias de la decisión (miscelánea) | 0,1% | 1,0 | 1,3 | | 0,1% | 1,7 | 1,4 | |
| Estadística, probabilidad e incertidumbre | 0,1% | 1,7 | 1,3 | | 0,1% | 2,9 | 1,3 | |
| Gestión de tecnología e innovación | 0,2% | 1,3 | 0,8 | | 0,1% | 0,9 | 0,8 | |
| Sistemas de información y gestión | 0,4% | 0,5 | 0,7 | *** | 0,2% | 0,4 | 0,3 | *** |
| Economía, econometría y finanzas | 1% | 1,5 | 1,7 | *** | 3% | 0,8 | 0,8 | |
| Economía y econometría | 0,5% | 1,4 | 1,1 | | 0,4% | 1,0 | 0,3 | *** |
| Economía, econometría y finanzas (miscelánea) | 0,2% | 1,9 | 1,3 | ** | 0,1% | 0,8 | 0,9 | |
| Finanzas | 0,2% | 0,7 | 0,8 | | 0,1% | 0,1 | -0,1 | *** |
| Negocios, gestión y contabilidad | 1% | 0,9 | 0,8 | ** | 1% | 0,9 | 0,4 | *** |
| Ciencias de la gestión e investigación operativa | 0,2% | 0,8 | 0,7 | | 0,2% | 0,5 | 0,1 | *** |
| Comportamiento organizacional y gestión de recursos humanos | 0,1% | 1,3 | 1,0 | | 0,1% | 0,6 | 0,5 | *** |
| Contabilidad | 0,1% | 0,9 | 0,9 | | 0,05% | 0,2 | 0,3 | *** |
| Estrategia y gestión | 0,4% | 0,8 | 0,6 | *** | 0,4% | 0,7 | 0,5 | *** |
| Gestión comercial e internacional | 0,3% | 0,6 | 0,5 | *** | 0,2% | 0,7 | 0,7 | * |
| Gestión de turismo, ocio y hostelería | 0,1% | 1,5 | 1,2 | | 0,1% | 1,1 | 1,3 | |
| Ingeniería industrial y de fabricación | 0,8% | 0,3 | 1,5 | *** | 0,8% | 0,2 | 0,3 | *** |
| Marketing | 0,1% | 0,3 | 0,9 | | 0,1% | 0,1 | 0,9 | |

| Grandes Áreas<br>Áreas<br>Disciplinas | Documentos | | | | Citas | | | |
|---|---|---|---|---|---|---|---|---|
| Negocios, gestión y contabilidad (miscelánea) | 0,2% | 0,8 | 0,6 | ** | 0,1% | 1,1 | 0,4 | |
| Sistemas de información gerencial | 0,1% | 0,1 | 0,8 | | 0,1% | 0,0 | 0,5 | |
| Otras ciencias sociales | 5% | 1,3 | 1,2 | *** | 1% | 0,5 | 0,7 | *** |
| Administración pública | 0,1% | 2,2 | 1,7 | *** | 0,05% | 2,0 | 1,6 | |
| Antropología | 0,1% | 1,2 | 0,9 | | 0,1% | 0,4 | 0,5 | *** |
| Arqueología | 0,1% | 1,4 | 1,1 | | 0,04% | 1,3 | 1,3 | |
| Bibliotecas y ciencias de la información | 0,1% | 1,4 | 1,1 | | 0,1% | 0,5 | 0,3 | *** |
| Ciencia política y relaciones internacionales | 0,2% | 1,6 | 1,6 | *** | 0,1% | 1,4 | 1,1 | |
| Ciencias sociales (miscelánea) | 0,3% | 1,4 | 1,3 | ** | 0,2% | 0,7 | 0,7 | ** |
| Comunicación | 0,2% | 1,1 | 0,6 | ** | 0,1% | 0,5 | 0,2 | *** |
| Demografía | 0,04% | 0,8 | 1,2 | | 0,02% | 0,2 | 1,5 | * |
| Desarrollo | 0,2% | 2,6 | 0,4 | * | 0,1% | 1,6 | 0,4 | ** |
| Educación | 0,7% | 1,7 | 2,5 | *** | 0,4% | 0,9 | 1,6 | ** |
| Estudios culturales | 0,3% | 1,8 | 3,2 | ** | 0,1% | 2,4 | 5,4 | ** |
| Estudios de curso y ciclo de vida | 0,04% | 0,6 | 1,3 | * | 0,03% | 0,4 | 0,4 | ** |
| Estudios de género | 0,1% | 2,4 | 1,2 | | 0,02% | 1,8 | 0,6 | ** |
| Estudios urbanos | 0,1% | 1,2 | 1,6 | * | 0,1% | 0,0 | 0,4 | *** |
| Factores humanos y ergonomía | 0,04% | 0,5 | 3,9 | *** | 0,03% | 0,1 | 4,8 | *** |
| Geografía, planificación y desarrollo | 0,5% | 1,4 | 0,7 | ** | 0,4% | 1,3 | 1,1 | |
| Investigación en seguridad | 0,1% | 0,9 | 0,9 | | 0,04% | 0,5 | 0,6 | * |
| Ley | 0,3% | 0,7 | 0,2 | ** | 0,1% | 0,7 | 0,2 | *** |
| Lingüísitica y lenguaje | 0,3% | 0,7 | 1,2 | | 0,1% | 0,0 | 0,6 | |
| Salud (ciencias sociales) | 0,2% | 1,2 | 0,9 | | 0,1% | 1,0 | 0,7 | |
| Sociología y ciencia política | 0,5% | 1,5 | 1,3 | ** | 0,3% | 1,0 | 0,7 | ** |
| Transporte | 0,1% | 0,7 | 0,6 | *** | 0,1% | 1,0 | 0,8 | |
| Psicología | 1% | 0,8 | 0,8 | *** | 1% | 0,5 | 0,5 | *** |
| Biología del desarrollo | 0,1% | 1,6 | 2,6 | *** | 0,1% | 0,9 | 1,4 | |
| Neuropsicología y psicología fisiológica | 0,1% | 0,6 | 0,8 | | 0,1% | 0,4 | 0,6 | |
| Psicología (miscelánea) | 0,2% | 1,3 | 1,1 | | 0,2% | 1,4 | 1,0 | |
| Psicología aplicada | 0,1% | 0,6 | 0,7 | *** | 0,1% | 0,2 | 0,3 | *** |
| Psicología clínica | 0,2% | 0,9 | 0,6 | *** | 0,2% | 0,5 | 1,0 | |
| Psicología cognitiva y experimental | 0,1% | 0,7 | 6,6 | ** | 0,1% | 0,3 | 6,5 | |

| Grandes Áreas<br>Áreas<br>Disciplinas | Documentos | | | | Citas | | | |
|---|---|---|---|---|---|---|---|---|
| Psicología social | 0,2% | 0,5 | 0,5 | *** | 0,1% | 0,1 | 0,2 | *** |
| Multidisciplinarias | 1% | 1,1 | 0,9 | | 2% | 1,2 | 1,3 | |
| Multidisciplinarias | 1% | 1,1 | 0,9 | | 2% | 1,2 | 1,3 | |
| Multidisciplinarias | 1,0% | 1,1 | 1,6 | * | 1,5% | 1,3 | 1,4 | |

Notas: % mundial corresponde a la proporción que representa cada disciplina en el total mundial de la producción científica; VCR trienio corresponde al valor calculado del índice VCR para el trienio 2016-2019; VCR regresión corresponde al valor proyectado 2019 según la modelización de la sección 2. *estadísticamente significativo al 10% **estadísticamente significativo al 5% ***estadísticamente significativo al 1%.